# Log-Periodic Power Law Singularities in Landslide Dynamics: Statistical Evidence from 52 Crises

Qinghua Lei[1], Didier Sornette[2]

[1]Department of Earth Sciences, Uppsala University, Sweden
[2] Institute of Risk Analysis, Prediction and Management, Academy for Advanced Interdisciplinary Studies, Southern University of Science and Technology, Shenzhen, China
Corresponding author: Qinghua Lei (qinghua.lei@geo.uu.se)

**Key points:**

- Landslide acceleration crises exhibit spontaneous power law finite-time singularities decorated by log-periodic oscillations.

- Parametric and nonparametric tests provide clear evidence of log-periodicity associated with discrete scale invariance in real landslides.

- Log-periodic landslide behavior may arise from stress drop and stress corrosion, alongside the interplay of inertia, damage, and healing.

**Abstract**

Landslide movements typically show a series of progressively shorter quiescent phases, punctuated by sudden bursts during an acceleration crisis. We propose that such intermittent rupture phenomena can be described by a log-periodic power law singularity model. Amounting mathematically to a generalization of the power law exponent from real to complex numbers, this model captures the partial break of continuous scale invariance to discrete scale invariance that is inherent to the intermittent dynamics of damage and rupture processes in heterogeneous geomaterials. By performing parametric and nonparametric tests on a large dataset of 52 landslides, we present empirical evidence and theoretical arguments demonstrating the statistical significance of log-periodic oscillations decorating power law finite-time singularities during landslide crises. Log-periodic landslide motions may stem from the interaction between frictional stress drop along geological structures and stress corrosion damage in rock bridges, as well as the interplay of inertia, damage, and healing.

**Plain Language Summary**

Forecasting landslides that threaten life and property remains a significant challenge. A major difficulty arises from the sporadic slope rupture behavior, typically characterized by a sequence of progressively shorter and shorter quiescent decelerating phases interspersed with sudden and intense accelerating bursts, rather than a smooth, monotonic progression of deformation and damage. This seemingly erratic pattern complicates landslide forecast and challenges conventional time-to-failure models that often assume a simple smooth power law acceleration. We propose a log-periodic power law singularity model, which can much better capture the intermittent, non-monotonic slope rupture dynamics at the site scale. We compile a large dataset of 52 landslides worldwide, including rockfalls, rockslides, clayslides, and embankment slopes, monitored by various instruments such as extensometers, reflectors, inclinometers, satellites, LiDAR, and SAR. We perform parametric and nonparametric tests on this dataset to provide clear evidence of log-periodic oscillations improving on the description in term of a power law acceleration during landslide crises. We reveal that log-periodicity in landslides may arise from a complex interplay of friction, damage, inertia, and healing in heterogeneous geomaterials. Our results have important implications for landslide forecast, because understanding and characterizing log-periodicity could transform intermittency from a hindrance into valuable information for improving predictions.

## 1 Introduction

Landslides occur across a wide range of Earth surface environments, posing severe threats to life and property (Froude & Petley, 2018; Lacroix et al., 2020; Pánek et al., 2024; Petley, 2012). Therefore, forecasting catastrophic slope failures is a fundamental goal of landslide hazard analysis. Over the past decades, great efforts have been dedicated to developing and deploying high-precision monitoring technologies to observe unstable slope movements (Casagli et al., 2023; Crosta et al., 2017), aiming to detect precursory features of imminent catastrophic failure events.

Several mechanisms and models have been proposed for understanding the approximate power law time-to-failure dynamics (Bell et al., 2011; Bufe & Varnes, 1993; Fukuzono, 1985; Kilburn & Petley, 2003; Voight, 1988, 1989) often observed in landslides. One mechanism builds on the analogy between catastrophic failure and critical points, based on the intrinsic scaling symmetry of power laws which is reminiscence of the scaling invariance symmetry enjoyed by critical phase transitions (Ausloos, 1986; Sornette, 2006). Other mechanisms have been proposed, such as the slider-block friction model, which attributes accelerating displacements to frictional instabilities along sliding surfaces (Handwerger et al., 2016; Helmstetter et al., 2004; Paul et al., 2024; Poli, 2017; Yamada et al., 2016) as well as stress corrosion-driven damage accumulation (Cornelius & Scott, 1993; Kilburn & Voight, 1998; Main, 2000; Sammis & Sornette, 2002). Modeling the time-to-failure dynamics as a finite-time power law singularity has been used to forecast the timing of landslide failures and develop early warning systems (Bell, 2018; Crosta & Agliardi, 2002, 2003; Intrieri et al., 2012, 2019; Leinauer et al., 2023). However, significant uncertainties have been found in applying the power law model for time-to-failure prediction, primarily due to the sporadic nature of slope rupture phenomena, which challenges the assumption of a smooth, monotonic power law acceleration.

The Log-Periodic Power Law Singularity (LPPLS) model, which incorporates log-periodic corrections to the power law trend, has been developed to capture the intermittent oscillations of heterogeneous systems approaching global breakdown (Anifrani et al., 1995; Sornette & Sammis, 1995; Johansen & Sornette, 1998). Recently, this model was shown to provide superior fits compared to conventional power law models, based on the thorough analysis of a comprehensive global dataset of historical geohazard events, including landslides, rockbursts, glacier breakoffs, and volcanic eruptions (Lei & Sornette, 2024). By leveraging the irregular and intermittent patterns in rupture dynamics, the LPPLS model transforms unsteady non-monotonous signals, traditionally perceived as noise, into essential components of the predictive framework, offering a promising tool for forecasting catastrophic events.

This study aims to provide further empirical evidence and theoretical arguments on the presence of log-periodicity in landslides, which is of both fundamental and practical interest. From a fundamental perspective, log-periodicity signals a spontaneous hierarchical organization of damage in heterogenous systems, offering insights into the mechanisms that drive rupture dynamics during landslide crises. From a practical perspective, log-periodicity can enhance the reliability of failure time forecast by "locking" the model fit into the accelerating oscillatory pattern of landslide motions. The rest of the Letter is organized as follows. Section 2 introduces the LPPLS model with the parametric calibration and nonparametric test methods described. Section 3 shows the results of our analysis based on a large dataset of 52 landslides with 100 time series. Finally, section 4 presents a discussion on the statistical significance of log-periodicity in landslides together with an interpretation of possible underlying mechanisms and implications for forecasting slope failures.

## 2 Methodology

The displacement behavior of a slope during an acceleration crisis is usually modeled by the following nonlinear dynamic equation (Crosta & Agliardi, 2002, 2003; Lei et al., 2023; Lei & Sornette, 2023; Voight, 1988, 1989):

$$\frac{d^2\Omega}{dt^2} = \eta \left(\frac{d\Omega}{dt}\right)^\alpha, \text{ with } \alpha > 1, \quad (1)$$

where $\Omega$ is displacement, $t$ is time, $\eta$ is a positive constant, and $\alpha$ is an exponent defining the degree of nonlinearity. The condition $\alpha > 1$ guarantees the existence of positive feedbacks (Main, 1999; Sammis & Sornette, 2002), leading to a super-exponential dynamic characterized by a finite-time singularity at a critical time $t_c$, around which an abrupt transition into a new regime would occur. Here, close to or beyond $t_c$, the system may either shift into an inertia-dominated regime of dynamic rupture or instead self-correct into a stabilized state. This singular behavior can be seen by integrating equation (1), yielding:

$$\frac{d\Omega}{dt} = \kappa(t_c - t)^{-\xi}, \text{ with } \xi > 0, \quad (2)$$

where $\kappa = (\xi/\eta)^\xi$, $\xi = 1/(\alpha-1)$, with $\xi > 0$ (for $\alpha > 1$) ensuring a singular behavior at $t = t_c$. A further integration of equation (2) leads to the so-called power law time-to-failure model (Bufe & Varnes, 1993; Main, 1999; Voight, 1988, 1989):

$$\Omega(t) = A + B(t_c - t)^m, \text{ with } m < 1, \quad (3)$$

where $A$ and $B = -\kappa/m$ are constants, and $m = 1-\xi = (\alpha-2)/(\alpha-1)$ is called the singularity exponent. Equation (3) can be proven to be the general solution for $\Omega$ when $m < 1$ ($\alpha > 1$), including the special case of $m = 0$ (Lei & Sornette, 2024). For $0 < m < 1$ ($\alpha > 2$), $d\Omega/dt$ diverges at $t_c$, but $\Omega$ converges to a finite value $A$; for $m < 0$ ($1 < \alpha < 2$), both $d\Omega/dt$ and $\Omega$ diverge at $t_c$. This power law relation enjoys the symmetry of continuous scale invariance, where scaling $t_c-t$ by an arbitrary factor $\lambda$ leads to a corresponding scaling of the observable (for $m \leq 0$) or of the difference of the observable to its final value $A$ (for $0 < m < 1$) by the factor $\lambda^m$ which is independent of $t_c-t$.

Let us now explore a generalized description in which the critical exponent is extended from real to complex values $m+i\omega$. Indeed, complex exponents are expected to generically emerge in systems with out-of-equilibrium dynamics and frozen disorders (Saleur & Sornette, 1996) and as solutions of general renormalization group equations of systems approaching critical points (Saleur et al., 1996a, 1996b). Exceptions include homogenous systems at equilibrium, which is not relevant to describe irreversible out-of-equilibrium landslide dynamics. The first-order Fourier expansion of the general solution of $\Omega$ yields the following LPPLS formula (Anifrani et al., 1995; Lei & Sornette, 2024; Sornette & Sammis, 1995):

$$\Omega(t) = A + \{B + C\cos[\omega\ln(t_c - t) - \phi]\}(t_c - t)^m, \text{ with } m < 1. \tag{4}$$

By introducing three additional parameters, i.e., a constant $C$, an angular log frequency $\omega$, and a phase shift $\phi$, equation (4) contains a log-periodic correction with a relative amplitude of $C/B$ (typically on the order of $10^{-1}$) to the power law trend with the pre-factor $B$. Here, the continuous scale invariance is partially broken into a discrete scale invariance (Saleur et al., 1996a; Sornette, 1998), where the observable obeys scale invariance under scaling of $t_c-t$ by specific factors that are integer powers of a specific fundamental scaling ratio $\lambda = \exp(2\pi/\omega) > 1$. The local maxima of the log-periodic term in the LPPLS formula occur at times converging to $t_c$ according to a geometric time series $\{t_1, t_2, \ldots, t_k, \ldots\}$ with $t_c-t_k = \lambda^{-k}\exp(\phi/\omega)$ and $k$ being an integer. This geometric time series is formed by time points where the argument of the cosine function in equation (4) is an integer multiple of $2\pi$. With this embedded discrete hierarchy of time scales, the LPPLS model can capture the intermittent rupture dynamics with a geometric increase in burst frequency on the approach to $t_c$, arising from the localized and threshold nature of rupture in heterogeneous materials (Johansen & Sornette, 2000; Sornette, 2002). Here, equation (4) includes only the first correction term, while higher-order terms with decreasing amplitudes also exist but are in general relatively less significant (Zhou & Sornette, 2002a).

We implement a stable and robust parametric calibration scheme, briefly described as follows (see Supplementary Text S1 for more details). First, the Lagrange regularization approach (Demos & Sornette, 2019) is employed to detect the onset time $t_0$ of an acceleration crisis, based on which the optimal time window is defined; here, the end of this time window is fixed either at the last available data point (if the crisis culminates in a catastrophic failure) or at the time stamp of peak velocity during the crisis (if the landslide self-stabilizes afterward). Then, the optimal parameter values for the LPPLS model are determined by minimizing the sum of the squares of the residuals, which quantifies the difference between the model and data (Filimonov & Sornette, 2013).

We further employ a nonparametric test to assess the presence of log-periodic oscillations. The following transformation (Johansen & Sornette, 2001) is applied to compute the normalized residual $\epsilon$ with the leading power law trend eliminated:

$$\epsilon(t) = \frac{\Omega(t) - A - B(t_c - t)^m}{C(t_c - t)^m}, \tag{5}$$

which should be a pure cosine function, i.e., $\cos(\omega\ln\tau - \phi)$, if equation (4) perfectly describes the data. Here, $\tau$ is the normalized time defined as $\tau = (t_c-t)/(t_c-t_0)$ with $t_0$ being the start of the time window over which the LPPLS fit is performed. Even if the data points are originally evenly sampled in the linear time scale, the data expressed as a function of the logarithmic time $\ln\tau$ is unevenly spaced. This makes standard fast Fourier methods ill-suited to our problem. We thus use the Lomb spectral method (Lomb, 1976) to identify oscillatory components on the logarithmic time scale $\ln\tau$. The Lomb method conducts a harmonic analysis by performing a local least-squares fit to data samples with sinusoids centered at each data point in the time series (Supplementary Text S2). The advantage of the Lomb periodogram is that it can deal with data with nonequidistant sampling, well suited to our problem of detecting periodicity on the logarithmic time scale to a singularity.

## 3 Results

The first case study is the Veslemannen landslide, composed of high-grade metamorphic rocks and situated on a north-facing slope in Romsdalen, western Norway. A ground-based interferometric synthetic-aperture radar system has been installed since October 2014 to continuously monitor this actively moving landslide (Kristensen et al., 2021). Major acceleration events, accompanied by substantial surface displacements and pronounced velocity spikes, were recorded in 2017, 2018, and 2019 (Figure 1). This instability complex was mainly active during summer/autumn months, likely due to rainwater infiltration into the slope through the thawed upper frost zone. Eventually, on September 5, 2019, ~54,000 m³ of unstable rock collapsed. Over the five-year monitoring period, the slope cumulatively displaced ~19 m and ~4 m in the upper and lower regions, respectively.

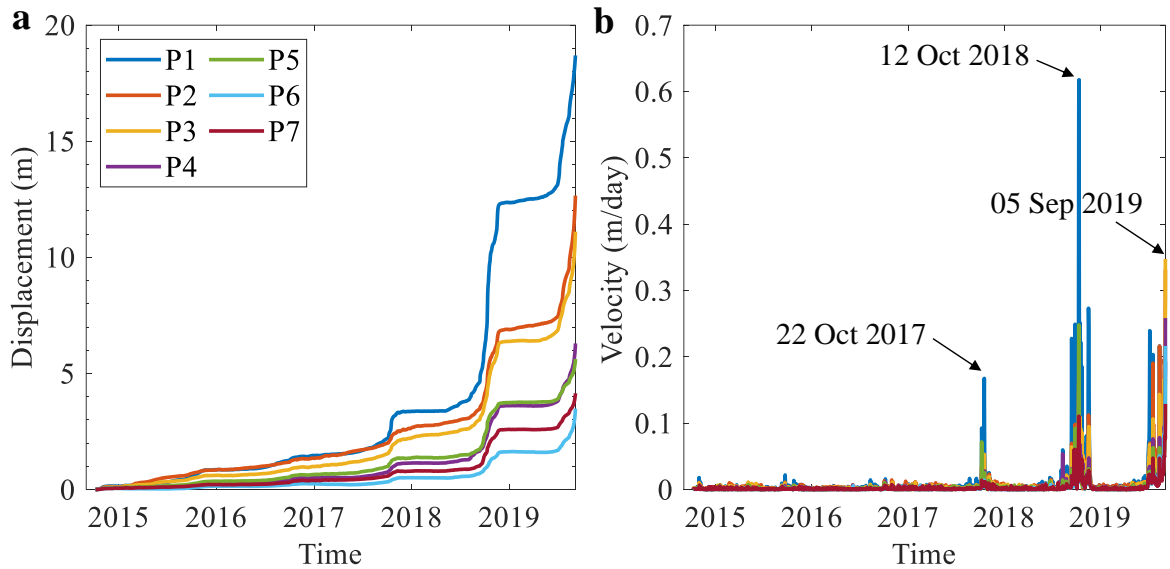

**Figure 1.** Monitoring data of the Veslemannen landslide, Norway. (a) Displacement time series recorded by seven radar points P1-P7. (b) Velocity time series with the peaks associated with the three major acceleration crises marked.

We perform the LPPLS calibration and Lomb periodogram analysis on the slope displacement time series during the three major acceleration crises in 2017, 2018, and 2019. This landslide exhibits a superimposition of acceleration and oscillations during all the three crises, as observed across all radar points (Figure 2 and Supplementary Figures 1-3). It is evident that the LPPLS model gives an excellent fit to the data, with the intermittency well captured. The duration of this log-periodic behavior increased progressively over the years. Log-periodic oscillations are evident in the $\epsilon$-$\ln\tau$ plot that closely follows a cosine function (insets of Figures 2a, 2c, and 2e), where some discrepancy in the small $\ln\tau$ region in Figure 2e may arise from the effects of higher-order harmonics of the oscillations. From the Lomb periodograms (Figures 2b, 2d, and 2f as well as Supplementary Figures 1-3 and Supplementary Table 2), we can clearly identify a dominant peak at the log frequency $f \approx 1.1$ (with the angular log frequency $\omega \approx 7.0$ and the scaling ratio $\lambda \approx 2.5$) for the 2017 and 2019 crises across all the radar points, while the dominant peak for the 2018 crisis occurs at $f \approx 0.8$ (with $\omega \approx 5.0$ and $\lambda \approx 3.5$). Interestingly, for the 2017 and 2018 crises, a harmonic can be respectively found at the log frequency of around 2.2-2.5 and 1.5-2.0 among most radar points (see Figures 2b, 2d, and 2f as well as Supplementary Figures 1-3), which corresponds roughly to the second harmonic $2f$. It is also evident that the maximum Lomb peak height $P_{\max}$ increases over time, from $17.4 \pm 1.9$ in 2017 to $22.8 \pm 2.1$ in 2018, and then to $31.6 \pm 6.1$ in 2019 (see Figures 2b, 2d, and 2f as well as Supplementary Figures 1-3 and Supplementary Table 2), confirming increasing statistical significance.

Following the same procedure, we perform the LPPLS calibration and Lomb periodogram analysis on various landslides based on a compiled global dataset of 52 landslides including 100 displacement time series (Supplementary Figures 4-14). This dataset covers different types of landslides including rockfalls, rockslides, clayslides, and embankment slopes, monitored by different instruments (e.g., extensometers, reflectors, distometers, inclinometers, satellites, LiDAR, and synthetic aperture radar) (Supplementary Table 1). Figure 3 shows typical examples analyzed.

Figure 3a presents the surface displacement of the Ruinon rockslide in Italy monitored by a distometer (Crosta & Agliardi, 2002, 2003). This rockslide consisting of 13 million m$^3$ phyllite exhibited significant episodic movements during 1997 to 2001, but no collapse occurred. This rockslide showed strong seasonal patterns in its displacement behavior, periodically accelerating during rainy seasons (summer and autumn) and then decelerating during dry seasons (winter and spring). However, after 2000, one can observe a clear log-periodic pattern, which is well described by the LPPLS model (Figure 3a, left). The presence of log-periodicity is also evident in the $\epsilon$-$\ln\tau$ plot (inset of Figure 3a, left), with the Lomb periodogram yielding $f = 2.15$, $\omega = 13.49$, and $\lambda = 1.59$ (Figure 3a, right).

Figure 3b shows the displacement data recorded by a bench mark on the La Clapière rockslide in France, which develops on a slope of metamorphic rocks mainly consisting of gneiss, amphibolites, and migmatites (El Bedoui et al., 2009). This rockslide experienced a major crisis between 1985 and 1987, before restabilising after late 1987 (Helmstetter et al., 2004). During this acceleration crisis, the landslide movements were found to correlate well with the flow rate of the Tinée river running along the slope toe (Sornette et al., 2004). The existence of log-periodicity is demonstrated by the LPPLS fit and the $\epsilon$-$\ln\tau$ pattern (though the amplitude of cyclical signals diminishes on the approach to $t_c$) (Figure 3b, left). On the Lomb periodogram, a peak is observed at $f = 1.02$ corresponding to $\omega = 6.41$ and $\lambda = 2.67$ (Figure 3b, right).

Figure 3c displays the data monitored by a Terrestrial LiDAR for the Puigcercós scarp in Spain, which experienced a major rockfall on December 3, 2013 (Royán et al., 2015). This rock face primarily consists of alternating layers of marl, sandstone, silt, and clay, capped by limestone. A clear log-periodic oscillating behavior decorating an overall power law acceleration can be seen in the displacement data which is well captured by the LPPLS model (Figure 3c). The $\epsilon$-$\ln\tau$ plot also confirms the presence of log-periodicity (inset of Figure 4a, left), and the Lomb spectral analysis indicates $f = 0.73$, $\omega = 4.58$, and $\lambda = 3.94$ (Figure 3c, right). Notably, the Lomb periodogram highlights a series of harmonics occurring at integer multiples of this fundamental frequency $f$—another signature of log-periodicity. The existence of log-periodicity is also found for other monitored areas of this scarp (Supplementary Figure 12).

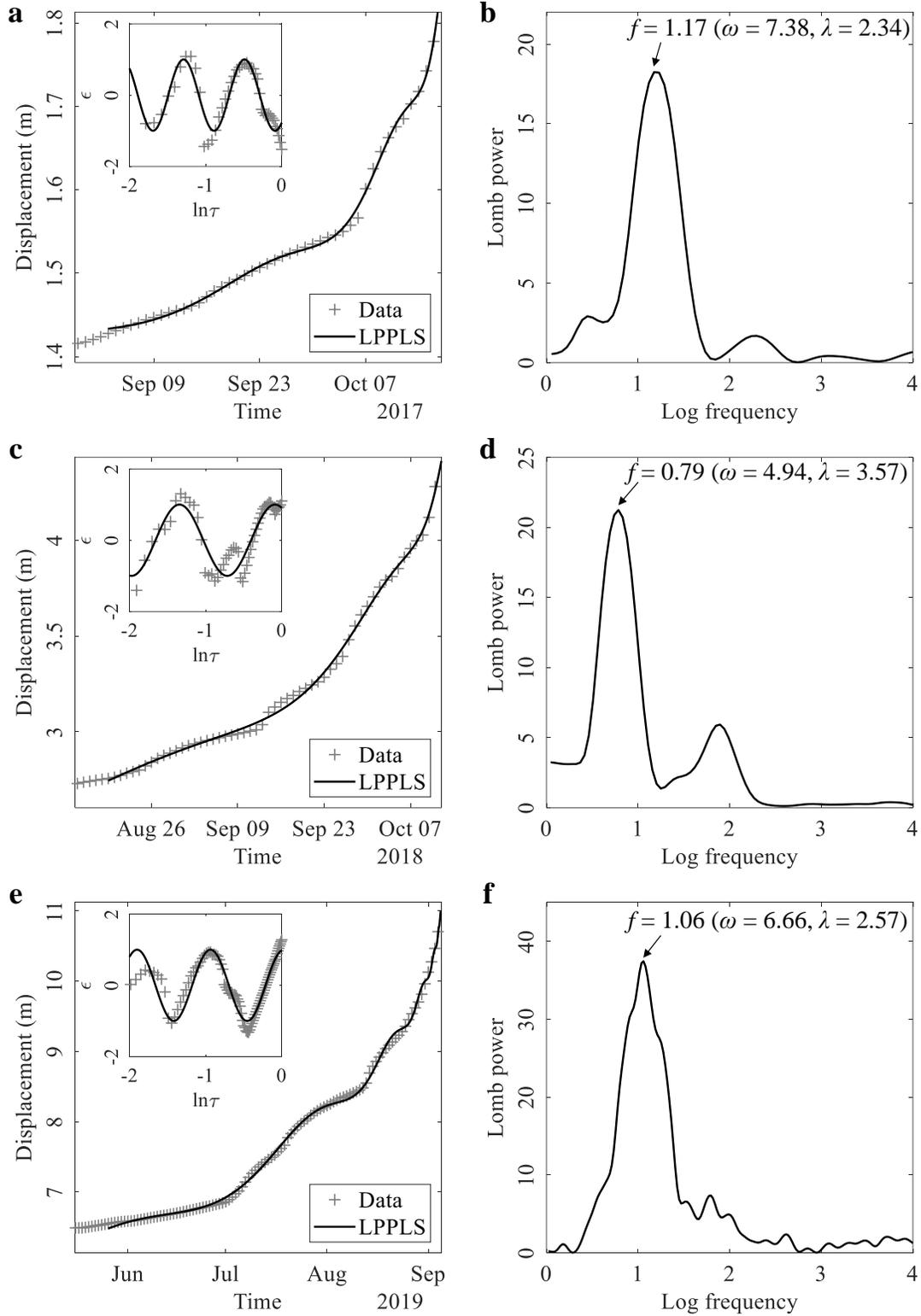

**Figure 2.** Time series of slope displacement of the Veslemannen landslide (recorded by radar point 3) fitted by the LPPLS model and the corresponding Lomb periodograms during the (a, b) 2017, (c, d) 2018, and (e, f) 2019 acceleration crises. Insets in the left panel show the normalized residual $\epsilon$, obtained from equation (5), as a function of the log normalized time $\tau = (t_c-t)/(t_c-t_0)$.

Figure 3d shows the displacement time series reconstructed based on synthetic-aperture radar images acquired by Sentinel-1 satellites for the Maoxian rockslide in China, which catastrophically failed on June 24, 2017 (Intrieri et al., 2018). This rockslide with a volume of ~13 million m$^3$ mainly consists of metamorphic sandstone, marbleized limestone, and phyllite (Fan et al., 2017). Before the final collapse, a pronounced accelerating oscillating behaviour is evident and well captured by the LPPLS model (Figure 3d, left). The existence of log-periodic oscillations is indicated by the sinusoidal-like signals in the $\epsilon$-ln$\tau$ plot (inset of Figure 3d, left) and the emergence of a major Lomb peak (Figure 3d, right), pointing to $f = 0.67$, $\omega = 4.23$, and $\lambda = 4.42$. Similar log-periodic characteristerics have been observed at two other measurement points (Supplementary Figures S8).

Figure 3e presents the displacement data recorded by a reflector installed on the Preonzo slope in Switzerland, where a volume of ~210,000 m$^3$ rock collapsed on May 15, 2012 (Gschwind et al., 2019; Loew et al., 2017). Similar displacement patterns are captured by other extensometers and reflectors instrumented on this rockslide (Supplementary Figures 10-11). This instability complex, predominantly composed of augen gneiss, exhibited a clear precursory acceleration phase, aligning well with the LPPLS model (Figure 3e and Supplementary Figures 10-11). Log-periodicity is evidenced by the cyclical pattern in the $\epsilon$-ln$\tau$ plot (inset of Figure 3e, left) and the Lomb periodogram (Figure 3e, right), which reveals $f = 0.67$, $\omega = 4.23$, and $\lambda = 4.42$, along with a series of harmonics.

Figure 3f displays the displacement time series of the Achoma landslide in Peru, derived from the images of high-frequency PlanetScope satellites (Lacroix et al., 2023). This landslide, situated in lacustrine deposits consisting of soils and weak rocks, experienced a catastrophic failure on June 18, 2020, prior to which a clear precursory accelerating motion was observed (Lacroix et al., 2023). Overall, the LPPLS model provides a good fit to the data (Figure 3f, left), though some scatter is evident, likely due to uncertainties associated with satellite-based measurements. However, log-periodicity is still identifiable in the $\epsilon$-ln$\tau$ plot (inset of Figure 3f, left) and the Lomb periodogram, revealing $f = 0.79$, $\omega = 4.99$, and $\lambda = 3.52$ (Figure 3f, right).

We compile and analyze the parameters derived from the Lomb analysis of 52 landslides, presenting histograms of selected key parameters in Figure 4 (see Supplementary Table S2 for the complete list). The angular log frequency $\omega$ ranges from 3 to 15, with a concentration around 5 (Figure 4a). Correspondingly, the scaling ratio $\lambda$ varies from 1.5 to 5, with a median value at around 3.5 (Figure 4b). Frequency distributions of $\omega$ and $\lambda$ derived from the Lomb method are in general compatible with those obtained from the LPPLS calibration (see Supplementary Figure 15; note that the discrepancy in the low $\omega$ and high $\lambda$ regions is attributed to the filter imposed for $\omega$ in the LPPLS calibration, as described in Supplementary Text 1). The histogram of maximum Lomb peak heights $P_{\max}$ of these landslides indicates that 94% exceed 5, 76% are beyond 10, 50% surpass 15, and 40% reach over 20 (Figure 4c), highlighting the significance of log-periodicity. The first-to-second peak ratio $\eta$, representing the ratio between the two highest peaks in each Lomb periodogram, shows that 75% of these landslides exceed 2 and 44% surpass 4 (Figure 4d), further strengthening the evidence of log-periodicity by the presence of both a fundamental log-frequency and its first harmonic.

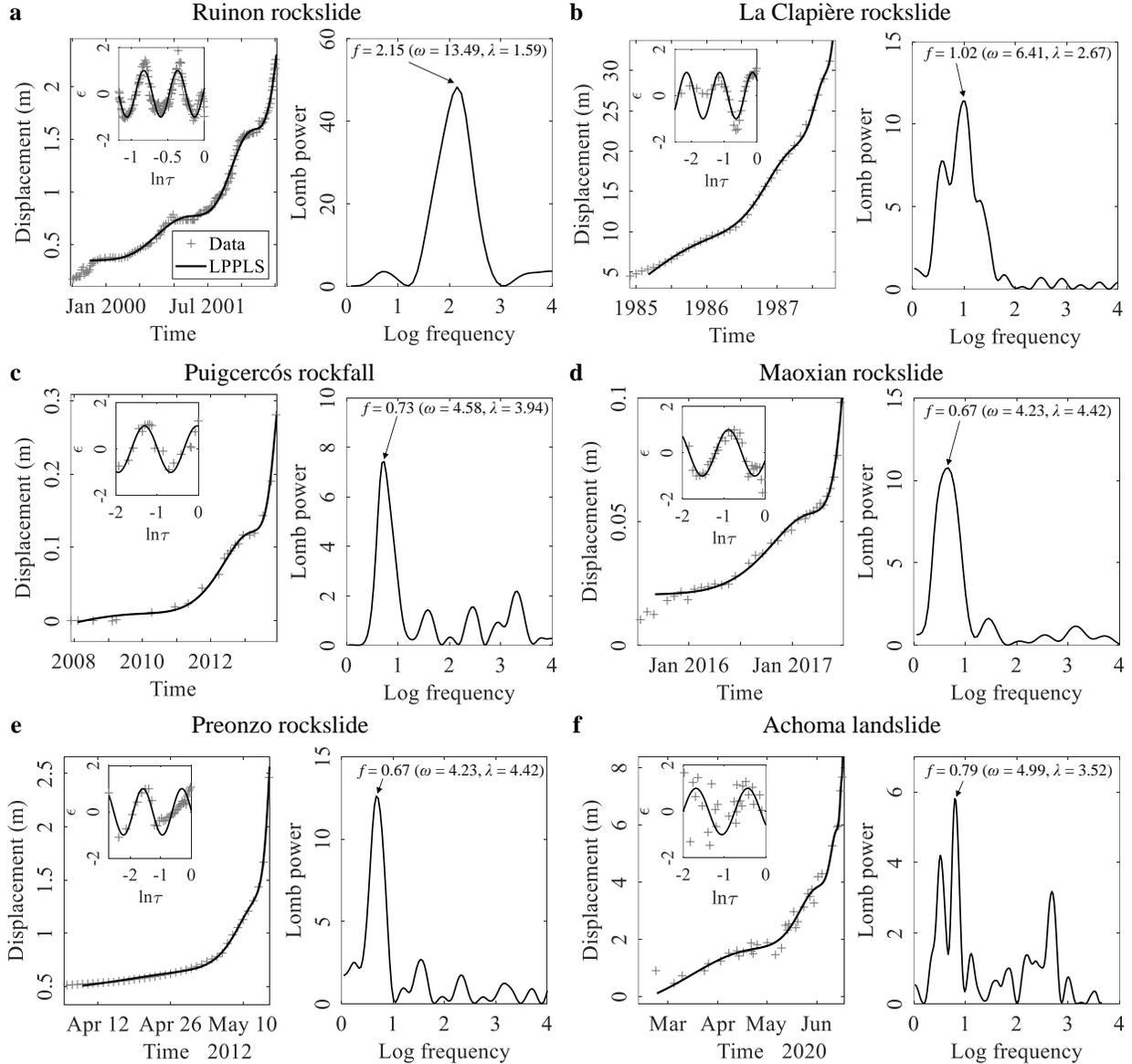

**Figure 3.** Time series of slope displacement data fitted by the LPPLS model with the corresponding Lomb periodogram analysis of (a) the Ruinon rockslide in Italy (based on data of distometer 7), (b) the La Clapière rockslide (based on data of bench mark 10) in France, (c) the Puigcercós rockfall (based on data of area 7) in Spain, (d) the Maoxian soilslide (based on data of measurement point 2) in China, (e) the Preonzo rockslide (based on data of reflector 2) in Switzerland, and (f) the Achoma landslide in Peru (based on data of PlanetScope satellites). Insets show the normalized residual $\epsilon$, obtained from equation (5), as a function of the log normalized time $\tau = (t_c-t)/(t_c-t_0)$.

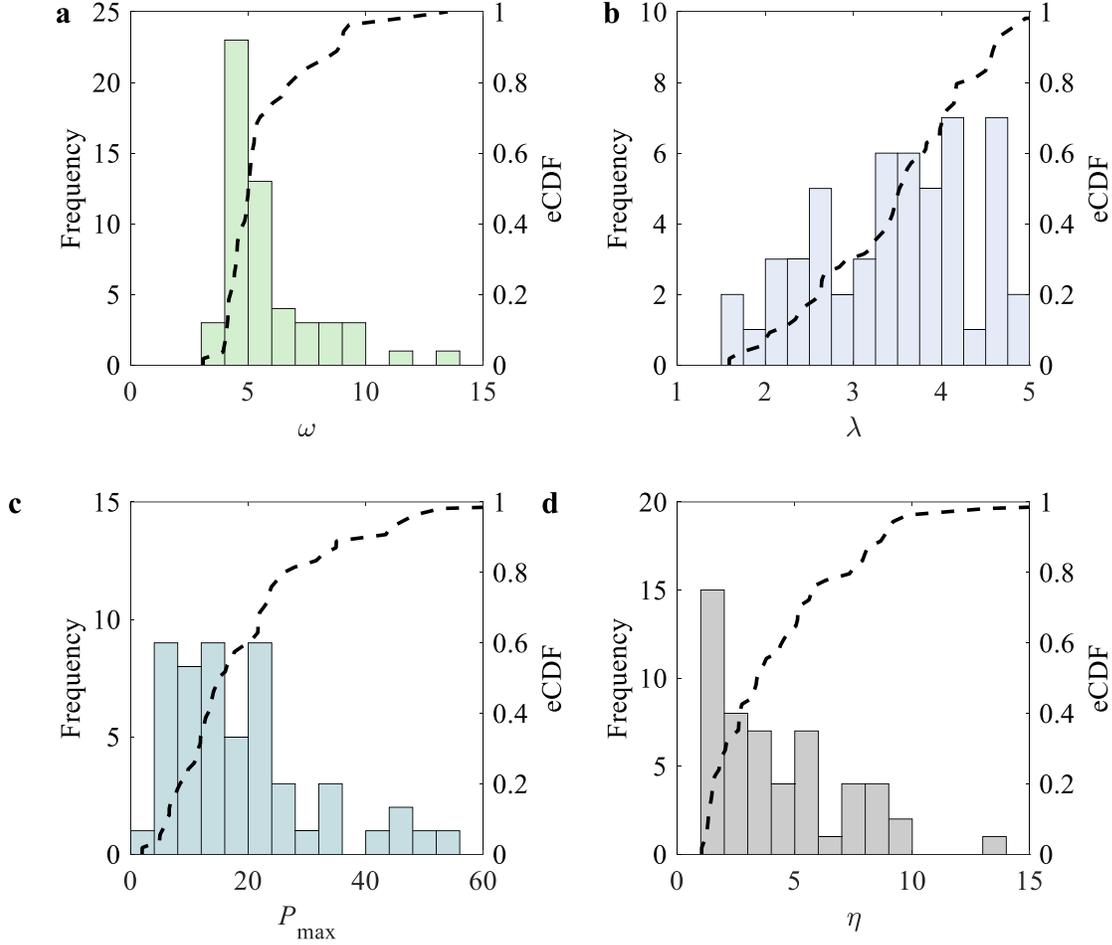

**Figure 4.** Histograms together with the empirical cumulative distribution function (eCDF) of the parameter values derived from the Lomb periodogram analysis of 52 landslides, including (a) the log-periodic frequency $\omega$, (b) the scaling ratio $\lambda$, (c) the maximum Lomb peak height $P_{max}$, and (d) the first-to-second peak ratio $\eta$.

## 4 Discussion

Our analysis suggests that log-periodicity seems to be a ubiquitous feature of landslide crises, indicated by the excellent LPPLS fit to the monitoring data of various landslides (Figures 2-3 and Supplementary Figures 1-14). Note that the possibility of overfitting by the LPPLS model has been ruled out based on Akaike and Bayesian information criteria in our previous study (Lei & Sornette, 2024). Parametric fitting reveals that the singularity exponent $m$, which characterizes the nonlinearity of power law acceleration, predominantly ranges from -1.5 to 0.5, with the corresponding exponent $\alpha = 1+1/(1-m)$ primarily varying between 1.4 and 3.0, consistent with previously reported $\alpha$ values for landslides (Intrieri et al., 2019). However, the $m$ values are found to concentrate around -0.5, with the corresponding $\alpha$ value near 1.7, which deviates from the commonly assumed $\alpha = 2.0$ in the inverse velocity method (Carlà et al., 2017; Fukuzono, 1985; Leinauer et al., 2023; Voight, 1988, 1989). The presence of log-periodic structures is qualified by the sinusoidal temporal evolution of residuals on a logarithmic time scale to the critical time, after removing the general power law trend, across almost all the landslides studied (see insets of Figures 2-3 and Supplementary Figures 1-14). Discrepancies observed in a few landslides stem from either data scarcity or the influence of harmonics not accounted for in the log-periodic formula (4), which expresses just the first-order correction in terms of the first log-periodic component to the pure power law; refer to Saleur et al. (1996a, 1996b) and Gluzman & Sornette (2002) for the general derivation of the full log-periodic solutions. The relative amplitude of log-periodic components is found to range between 0.005 and 0.15, with a concentration around 0.05, aligning with the conjecture of being on the order of $10^{-1}$ for systems undergoing failures (Sornette, 1998). Our nonparametric tests provide further evidence of log-periodicity in landslide crises. For example, the consistently high Lomb peaks for various landslides, with the majority exceeding 10 (Figure 4c), highlight the statistical significance of log-periodic components (Zhou & Sornette, 2002b). If the data were due to Gaussian noise, we can analytically derive the false-alarm probability quantifying the likelihood that random noise is incorrectly identified as a valid log-periodic signal (Supplementary Text S2). Out of the 100 time series analyzed, only 7 exhibit a false-alarm probability greater than 0.05 (Supplementary Table S2). It is possible that the noise in the data does not follow a Gaussian distribution; however, the high first-to-second Lomb peak ratios, with the majority exceeding 2 and about half surpassing 4, still offer strong evidence of log-periodicity (Zhou & Sornette, 2002b). This is further supported by the high signal-to-noise ratios—a metric broadly applicable to various types of noise (Zhou & Sornette, 2002a, 2002b)—for most landslides (Supplementary Table S2), with 96% greater than 1, 76% exceeding 1.5, and 58% surpassing 2.

The log-periodic characteristics offer valuable insights into the underlying mechanisms driving the intermittent rupture behavior of geomaterial masses during landslide crises. It reflects the presence of discrete scale invariance (Saleur et al., 1996a; Sornette, 1998), which is associated with complex critical exponents typically arising in nonunitary (dissipative) systems with out-of-equilibrium dynamics and quenched disorder (Saleur & Sornette, 1996). One possible mechanism for discrete scale invariance is the cascade of ultraviolet Mullins-Sekerka instabilities during crack growth, where larger cracks are less affected by screening and propagate faster, while smaller cracks are suppressed due to stress shadowing and crack interactions (Huang et al., 1997). This theory developed for a regular array of pre-existing cracks, supported by geological evidence found in natural outcrops (Ouillon, Sornette, et al., 1996), predicts $\lambda = 2$, which lies at the lower end of the range of $\lambda$ values reported in our current study. The variability of $\lambda$ values found here may reflect the presence of a heterogenous spatial and length distribution of pre-existing cracks and sliding surfaces. Another possible mechanism is the interplay between stress drop associated with rupture dynamics and stress corrosion during inter-rupture phases, as demonstrated in sandpile models with a mean-field prediction of $\lambda = 3.6$ (Lee & Sornette, 2000), aligning closely with our median $\lambda$ value of ~3.5 (Figure 4b). Log-periodicity could also arise from the interplay of inertia, damage, and healing (Ide & Sornette, 2002; Sornette & Ide, 2003), which may account for the significant variability in $\lambda$ values observed for different landslides, due to the possible variations in their healing properties. Furthermore, the interplay of heterogeneities and stress concentrations may also produce discrete scale invariance (Sahimi & Arbabi, 1996), while the presence of hierarchical fracture networks in geological media (Bonnet et al., 2001; Lei & Wang, 2016; Ouillon, Castaing, et al., 1996) might contribute as well, though discrete scale invariance can arise spontaneously in the absence of predefined hierarchical structures (Sornette, 1998). It is likely that these different mechanisms coexist and interact in real landslides, with the dominant mechanism varying both across different sites and over time for the same site. For instance, the Veslemannen landslide exhibits $\lambda \approx 2.5$ during the 2017 and 2019 crises, indicating a possible dominance of crack growth and interaction, while $\lambda \approx 3.5$ during the 2018 crisis may suggest a dominant interaction between frictional stress drop along geological structures and stress corrosion damage in rock bridges.

In addition to shedding light on landslide mechanisms, log-periodic signals could be practically useful for forecasting impending slope failures. More specifically, by "locking" into the oscillatory structure of rupture dynamics, time-to-failure predictions employing the LPPLS model can transform intermittency—traditionally viewed as a nuisance—into valuable information, thereby enhancing the precision of critical time of failure estimations (Lei & Sornette, 2024; Sornette, 2002). Its potential for prospective forecasting will be explored in our future work. Moreover, log-periodic signatures could serve as indicators for early warning, a concept proven effective for forecasting financial crises (Demirer et al., 2019; Zhang et al., 2016) and to be explored for landslides in the future. Our analysis of the Veslemannen landslide suggests that the significance of log-periodic oscillations tends to increase as rupture approaches, with Lomb peak heights rising over the years (see Figure 2 and Supplementary Figures 1-3). However, the Lomb nonparametric analysis may be less robust for prediction than the LPPLS parametric fit (Zhou & Sornette, 2002a).

## Acknowledgments

Q.L. acknowledges the National Academic Infrastructure for Supercomputing in Sweden (NAISS), partially funded by the Swedish Research Council through grant agreement no. 2022-06725, for awarding this project access to the LUMI supercomputer, owned by the EuroHPC Joint Undertaking and hosted by CSC (Finland) and the LUMI consortium. D.S. acknowledges partial support from the National Natural Science Foundation of China (Grant No. U2039202, T2350710802), from the Shenzhen Science and Technology Innovation Commission (Grant No. GJHZ20210705141805017) and the Center for Computational Science and Engineering at the Southern University of Science and Technology. We are grateful to Agnes Helmstetter and Jérôme Faillettaz for sharing the relevant monitoring data of the La Clapière and Séchilienne landslides, respectively.

Supporting Information for

**Log-Periodic Power Law Singularities in Landslide Dynamics:
Statistical Evidence from 52 Crises**

Qinghua Lei[1], Didier Sornette[2]

[1]Department of Earth Sciences, Uppsala University, Sweden
[2]Institute of Risk Analysis, Prediction and Management, Academy for Advanced Interdisciplinary Studies, Southern University of Science and Technology, Shenzhen, China

**Contents of this file**

Text S1 to S3
Figures S1 to S15
Tables S1 to S2

**Introduction**

This document provides supporting information to complement the methods, results, and discussions in the main Letter. Text S1 gives a detailed description of the calibration scheme for the LPPLS model. Text S2 shows the mathematical formulation of the Lomb method for detecting oscillatory components in unevenly sampled data. Text S3 describes the data acquisition approach in this study. Figures S1-S3 show LPPLS fits to the Veslemannen landslide displacement data (radar points 1, 2, 4, 5, 6, and 7) together with the corresponding Lomb periodograms during the 2017, 2018, and 2019 acceleration crises. Figures S4-S14 give the LPPLS fits and the corresponding Lomb periodograms for various other landslides during their acceleration crises. Figure S15 shows the histograms and the empirical cumulative distribution functions of the LPPLS parameters obtained for 52 landslides. Table S1 summarizes the key information (e.g. location, type, material, and so on) of the 52 landslides analyzed in the current study. Table S2 lists the LPPLS and Lomb parameters obtained for these landslides.

**Text S1.** Calibration scheme for the LPPLS model

Let us consider the time series data of slope displacement $\mathbf{\Omega} = \{\Omega_1, \Omega_2, ..., \Omega_N\}$ recorded at time $\mathbf{t} = \{t_1, t_2, ..., t_N\} \in [t_0, t_f]$, where $N$ is the total number of time stamps, and $t_0$ and $t_f$ respectively denote the start time and final time of the time window for model calibration. The LPPLS model is calibrated against the time series data using a robust calibration scheme (Filimonov & Sornette, 2013), briefly summarized as follows.

The original LPPLS formula is given as:

$$\Omega(t) = A + \{B + C\cos[\omega \ln(t_c - t) - \phi]\}(t_c - t)^m, \tag{S1}$$

whose parameter set $\mathbf{\theta} = \{A, B, C, t_c, m, \omega, \phi\}$ has seven parameters, with the former three being linear and the latter four being nonlinear. By defining $C_1 = C\cos\phi$ and $C_2 = C\sin\phi$, we can rewrite equation (S1) as:

$$\Omega(t) = A + B(t_c - t)^m + C_1(t_c - t)^m \cos[\omega \ln(t_c - t)] + C_2(t_c - t)^m \sin[\omega \ln(t_c - t)], \tag{S2}$$

where the new parameter set $\mathbf{\theta} = \{A, B, C_1, C_2, t_c, m, \omega\}$ still has seven parameters, but with the first four being linear and the last three being nonlinear. To determine these parameters, we define the cost function as the total sum of squared errors:

$$F(\mathbf{\theta}_{LPPLS}; \mathbf{\Omega}, \mathbf{t}) = \sum_{i=1}^{N} \varepsilon_i^2, \tag{S3}$$

where each residual is given by:

$$\varepsilon_i = \Omega_i - A - B(t_c - t_i)^m - C_1(t_c - t_i)^m \cos[\omega \ln(t_c - t_i)] - C_2(t_c - t_i)^m \sin[\omega \ln(t_c - t_i)] \tag{S4}$$

We minimize the cost function (S3) based on the ordinary least squares method to estimate the model parameters as:

$$\hat{\mathbf{\theta}} = \arg\min_{\mathbf{\theta}} F(\mathbf{\theta}; \mathbf{\Omega}, \mathbf{t}). \tag{S5}$$

By enslaving the four linear parameters $\{A, B, C_1, C_2\}$ to the three nonlinear ones $\{t_c, m, \omega\}$, the minimization problem reduces to:

$$\{\hat{t}_c, \hat{m}, \hat{\omega}\} = \arg\min_{t_c, m, \omega} F_1(t_c, m, \omega), \tag{S6}$$

where the profiled cost function is defined as:

$$F_1(t_c, m, \omega) = \min_{A, B, C_1, C_2} F(A, B, C_1, C_2, t_c, m, \omega) = F(t_c, m, \omega, \hat{A}, \hat{B}, \hat{C}_1, \hat{C}_2). \tag{S7}$$

The estimates for the four linear parameters $\{A, B, C_1, C_2\}$ can be obtained by solving the optimization problem for fixed values of the nonlinear parameters $\{t_c, m, \omega\}$:

$$\{\hat{A}, \hat{B}, \hat{C}_1, \hat{C}_2\} = \arg\min_{A, B, C_1, C_2} F(A, B, C_1, C_2, t_c, m, \omega), \tag{S8}$$

whose solution can be analytically obtained by solving the following system of linear equations:

$$\begin{bmatrix} N & \sum f_i & \sum g_i & \sum h_i \\ \sum f_i & \sum f_i^2 & \sum f_i g_i & \sum f_i h_i \\ \sum g_i & \sum f_i g_i & \sum g_i^2 & \sum g_i h_i \\ \sum h_i & \sum f_i h_i & \sum g_i h_i & \sum h_i^2 \end{bmatrix} \begin{bmatrix} \hat{A} \\ \hat{B} \\ \hat{C}_1 \\ \hat{C}_2 \end{bmatrix} = \begin{bmatrix} \sum \Omega_i \\ \sum \Omega_i f_i \\ \sum \Omega_i g_i \\ \sum \Omega_i h_i \end{bmatrix}, \tag{S9}$$

where $f_i = (t_c - t_i)^m$, $g_i = (t_c - t_i)^m \cos[\omega \ln(t_c - t_i)]$, and $h_i = (t_c - t_i)^m \sin[\omega \ln(t_c - t_i)]$.

We can further reformulate the minimization problem (S6) as:

$$\hat{t}_c = \arg\min_{t_c} F_2(t_c), \tag{S10}$$

where the cost function is written as:

$$F_2(t_c) = \min_{m,\omega} F_1(t_c, m, \omega) = F_1(t_c, \hat{m}, \hat{\omega}), \tag{S11}$$

such that the estimates for parameters $\{m, \omega\}$ can be obtained by solving:

$$\{\hat{m}, \hat{\omega}\} = \arg\min_{m,\omega} F_1(t_c, m, \omega). \tag{S12}$$

We impose a filter of $4.94 \leq \omega \leq 15$, with the lower and upper bounds preventing chaotic scenarios (Saleur et al., 1996b) and spurious oscillations (Filimonov & Sornette, 2013), respectively.

For a fixed end time $t_f$, the optimal start time $t_0$ of the time window for the LPPLS model calibration can be determined using the Lagrange regularization approach (Demos & Sornette, 2019) minimizing the following cost function:

$$\tilde{F}'(t_0) = \tilde{F}(t_0) - \chi N(t_0), \tag{S13}$$

where $\chi$ is the Lagrange parameter, and $\tilde{F}(t_0)$ is the normalized sum of squared residuals:

$$\tilde{F}(t_0) = \frac{F}{N(t_0) - n}, \tag{S14}$$

where $F$ is given by equation (S3) and $n$ is the number of degrees of freedom of the model (i.e., 7 for the LPPLS model). The Lagrange parameter $\chi$ may be estimated through a linear regression of $\tilde{F}(t_0)$ versus $t_0$.

**Text S2.** Lomb method

Assume a time series of signal $y(x_i)$ unevenly sampled at times $x_i$, where $i = 1, 2, \ldots, N$. The mean and variance of the data can be computed as:

$$\bar{y} = \frac{1}{N} \sum_{1}^{N} y_i, \tag{S15}$$

and

$$\sigma^2 = \frac{1}{N-1} \sum_{1}^{N} (y_i - \bar{y})^2. \tag{S16}$$

The normalized Lomb periodogram is then calculated as (Zhou & Sornette, 2002b):

$$P(\omega) = \frac{1}{2\sigma^2} \left\{ \frac{\left[\sum_{1}^{N}(y_i - \bar{y})\cos\omega(x_i - \xi)\right]^2}{\sum_{1}^{N} \cos^2\omega(x_i - \xi)} + \frac{\left[\sum_{1}^{N}(y_i - \bar{y})\sin\omega(x_i - \xi)\right]^2}{\sum_{1}^{N} \sin^2\omega(x_i - \xi)} \right\}. \tag{S17}$$

where $\omega = 2\pi f$ is the angular frequency with $f$ being the frequency, and $\xi$ is given by:

$$\tan(2\omega\xi) = \left(\sum_{1}^{N} \sin 2\omega x_i\right) \bigg/ \left(\sum_{1}^{N} \cos 2\omega x_i\right). \tag{S18}$$

In the Lomb periodogram, peaks at specific angular frequencies indicate the potential presence of periodic components. The highest peaks correspond to the dominant frequencies in the time series. The higher the peak, the greater the statistical significance of the corresponding periodic component. If the data are characterized by Gaussian noise (independently normally distributed), the probability for a detected peak to exceed a given height $z$, can be derived as:

$$p(>z) = 1 - (1 - e^{-z})^M, \tag{S19}$$

where $M$ is the effective number of independent frequencies. Zhou & Sornette (2002b) have generalized this expression for a large set of noise properties beyond Gaussian, with power law tails as well as long-range dependence.

Our analysis is aimed at detecting log-periodicity in the signal at a function of time. This amounts to detect periodicity of the normalized residual $\epsilon$ defined by expression (S4) as a function of normalised logarithmic time $\ln\tau$, where $\tau = (t_c - t)/(t_c - t_0)$ with $t_0$ being the start of the time window over which the LPPLS fit is performed. We thus substitute $x_i = \ln\tau_i$ and $y_i = \epsilon_i$ into equation (S17) to obtain the corresponding Lomb spectrum. Then, the maximum peak height $P_{max}$ can be obtained as the maximum value of the Lomb periodogram $P(\omega)$, with the correponding dominant (log-)frequency $f_{Lomb}$ or angular (log-)frequency $\omega_{Lomb}$ identified. It is important to keep in mind that, given the definition of the reduced variable $\ln\tau = \ln[(t_c - t)/(t_c - t_0)]$, $f_{Lomb}$ and $\omega_{Lomb}$ are not conventional frequencies with dimensions of inverse of time. They are dimensionless and encode the existence of discrete scale invariance with a preferred scaling ratio $\lambda_\omega = \exp(2\pi/\omega) > 1$, as explained in the Letter.

Further, the false-alarm probability $p_{FA}$ for the highest peak can be obtained by substituting the maximum peak height $P_{max}$ into equation (S19) with $M = N$. In addition, the ratio of first-to-second highest peak $\eta$ can indicate the relative significance of the highest Lomb peak (Zhou & Sornette, 2002b). The signal-to-noise ratio $\gamma$ can also be estimated as:

$$\gamma = \left( \frac{4P_{max}}{N - 2P_{max}} \right)^{1/2}, \tag{S20}$$

which approximately holds for different types of noises, including non-Gaussian noise (Zhou & Sornette, 2002a, 2002b).

**Text S3.** Data acquisition approach

In this study, we have compiled a large dataset of 52 landslides (see Supplementary Table S1). These data were collected through two primary methods: (1) exported directly from the monitoring system and obtained from either published dataset/database or from the authors (indicated as "Original" in Supplementary Table S1), and (2) digitized from figures in published literature using digitization software (indicated as "Digitized" in Supplementary Table S1). For most of digitized data, we employ the software PlotDigitizer Pro (https://plotdigitizer.com) to retrieve the data from the published literature. The references for all the data are indicated in Supplementary Table S1.

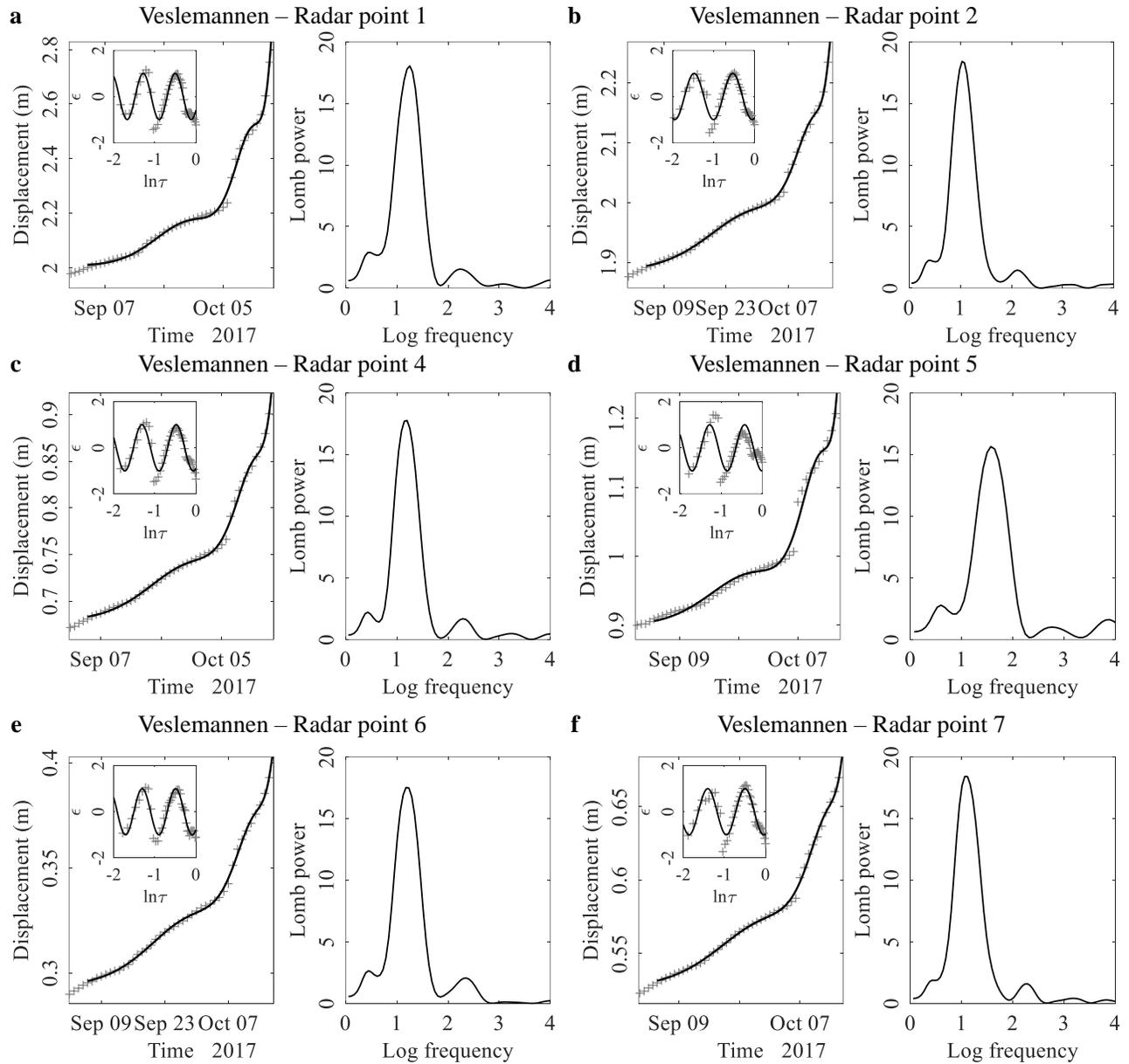

**Figure S1.** Time series of the displacement data of the Veslemannen landslide fitted to the LPPLS model (insets show normalized residual $\epsilon$ as a function of log normalized time $\tau$) and the corresponding Lomb periodograms during the 2017 acceleration crisis.

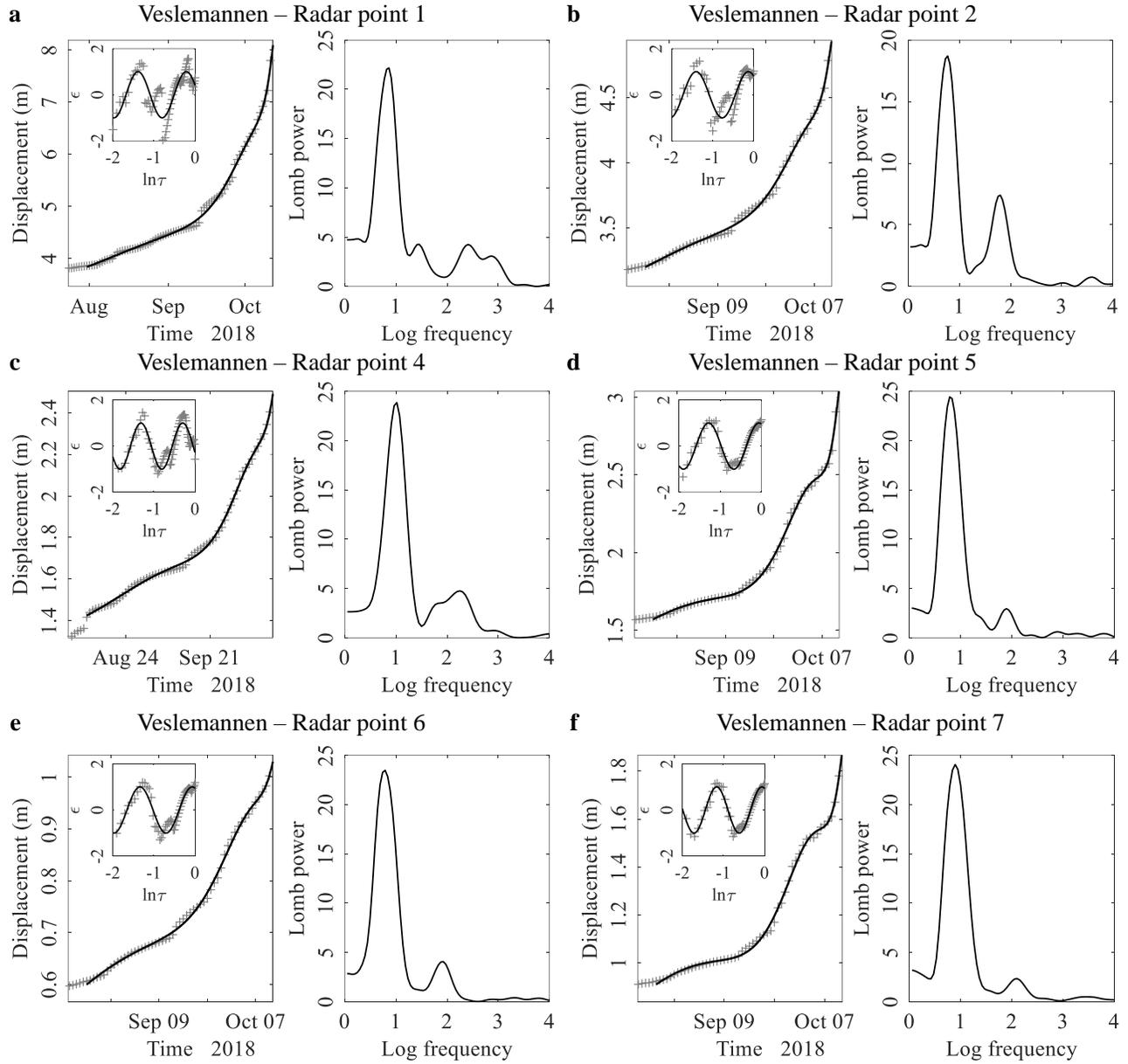

**Figure S2.** Time series of the displacement data of the Veslemannen landslide fitted to the LPPLS model (insets show normalized residual $\epsilon$ as a function of log normalized time $\tau$) and the corresponding Lomb periodograms during the 2018 acceleration crisis.

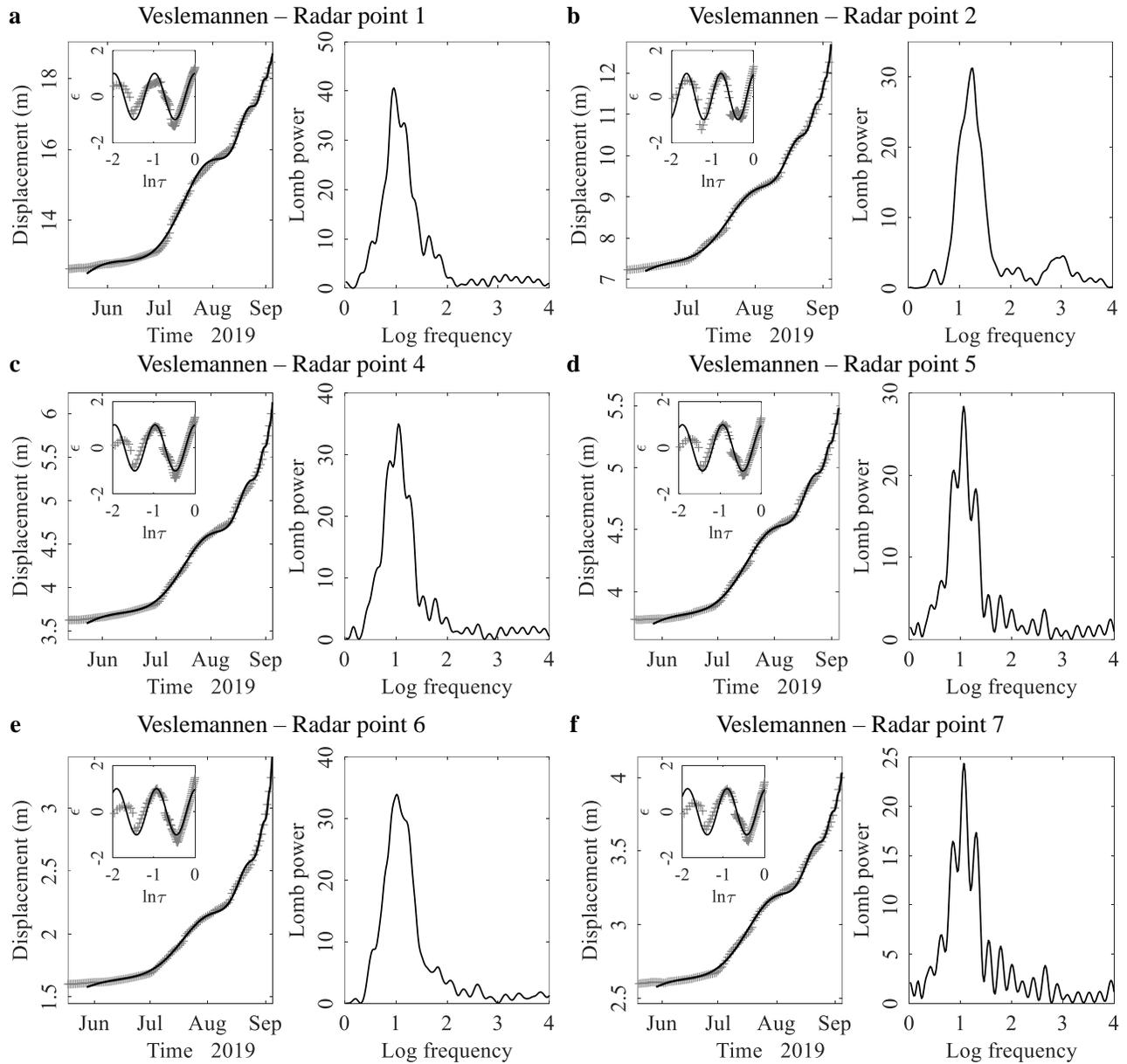

**Figure S3.** Time series of the displacement data of the Veslemannen landslide fitted to the LPPLS model (insets show normalized residual $\epsilon$ as a function of log normalized time $\tau$) and the corresponding Lomb periodograms during the 2019 acceleration crisis.

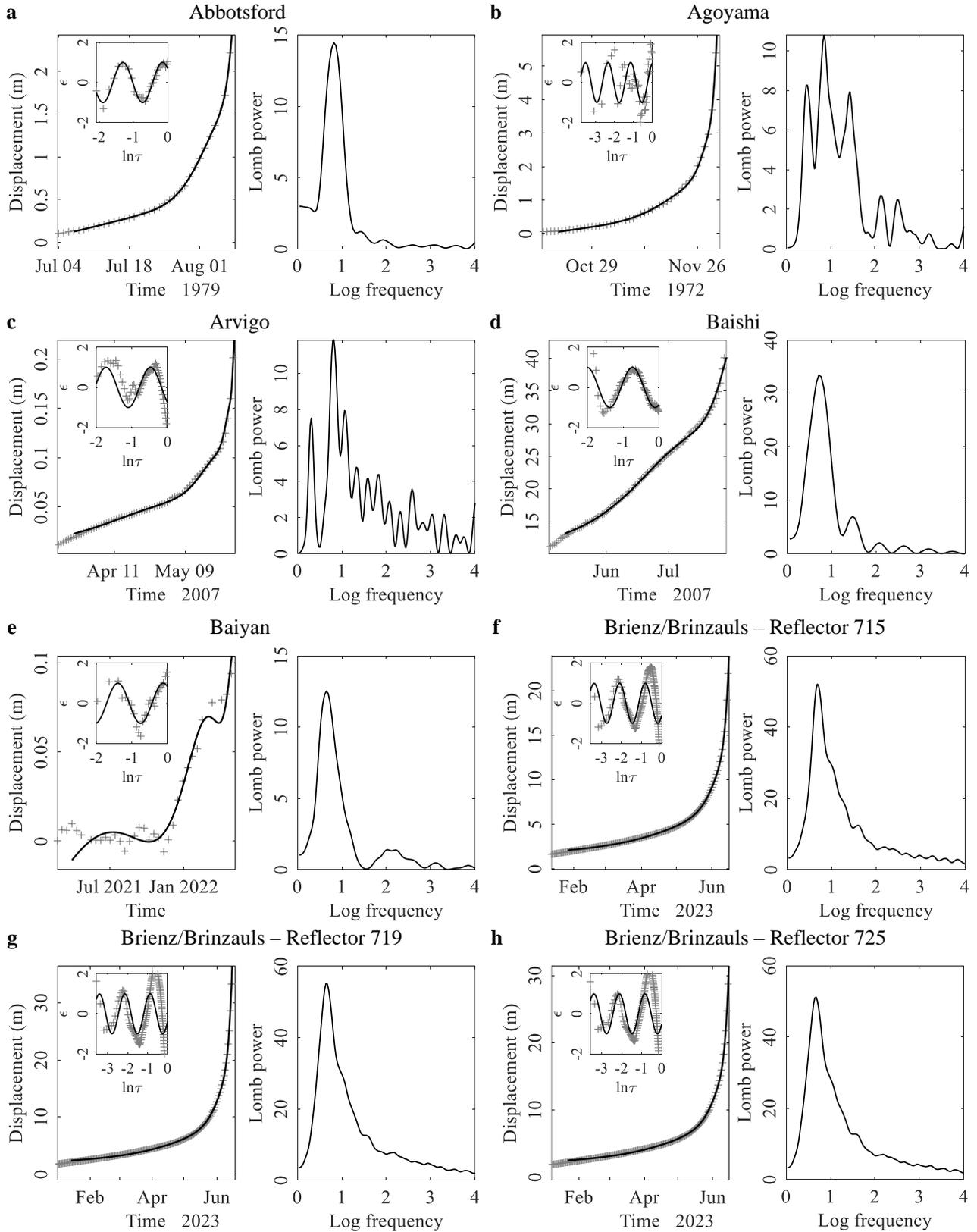

**Figure S4.** Time series of the displacement data of various landslides fitted to the LPPLS model (insets show normalized residual $\epsilon$ as a function of log normalized time $\tau$) and Lomb periodograms during acceleration crises.

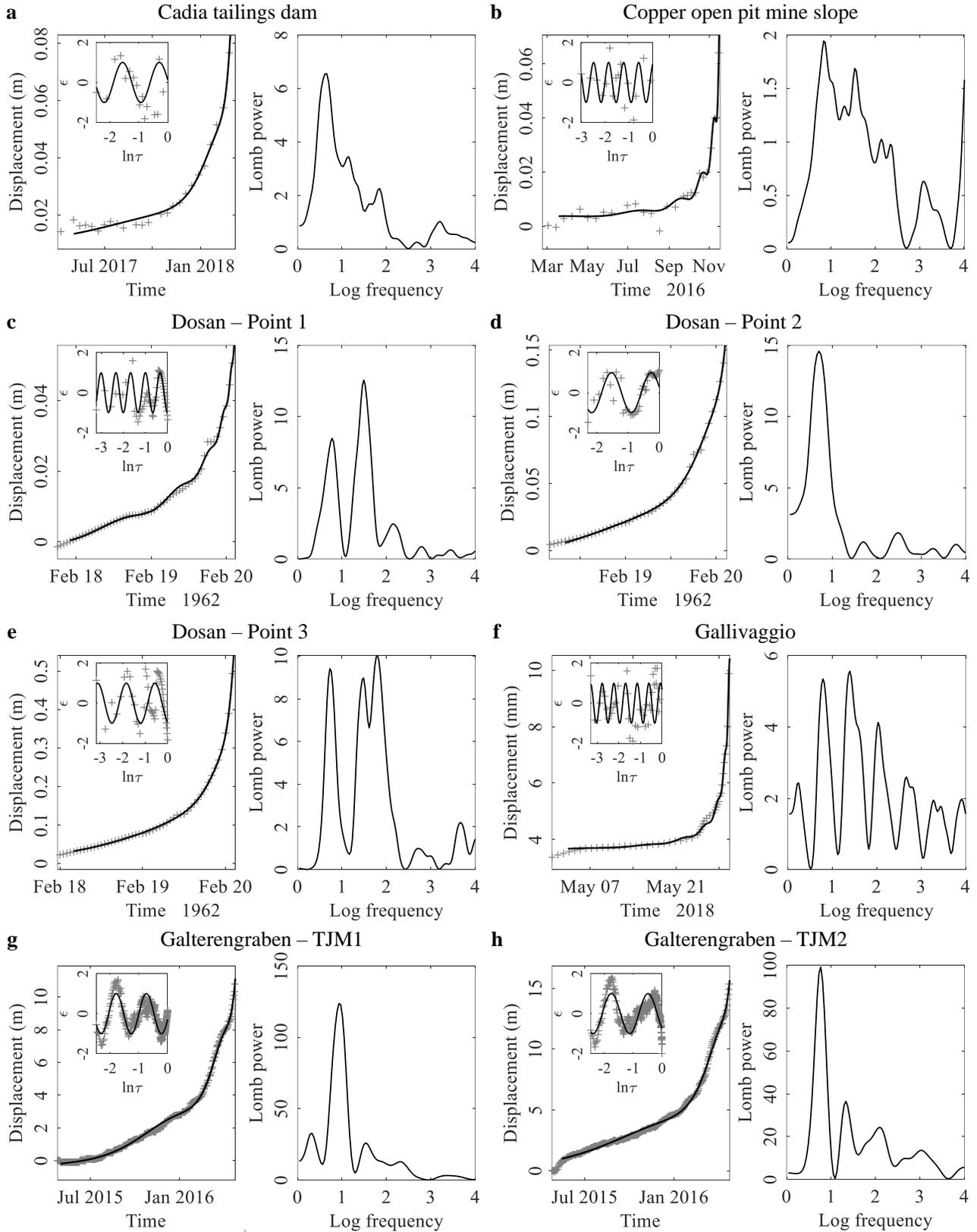

**Figure S5.** Time series of the displacement data of various landslides fitted to the LPPLS model (insets show normalized residual $\epsilon$ as a function of log normalized time $\tau$) and Lomb periodograms during acceleration crises.

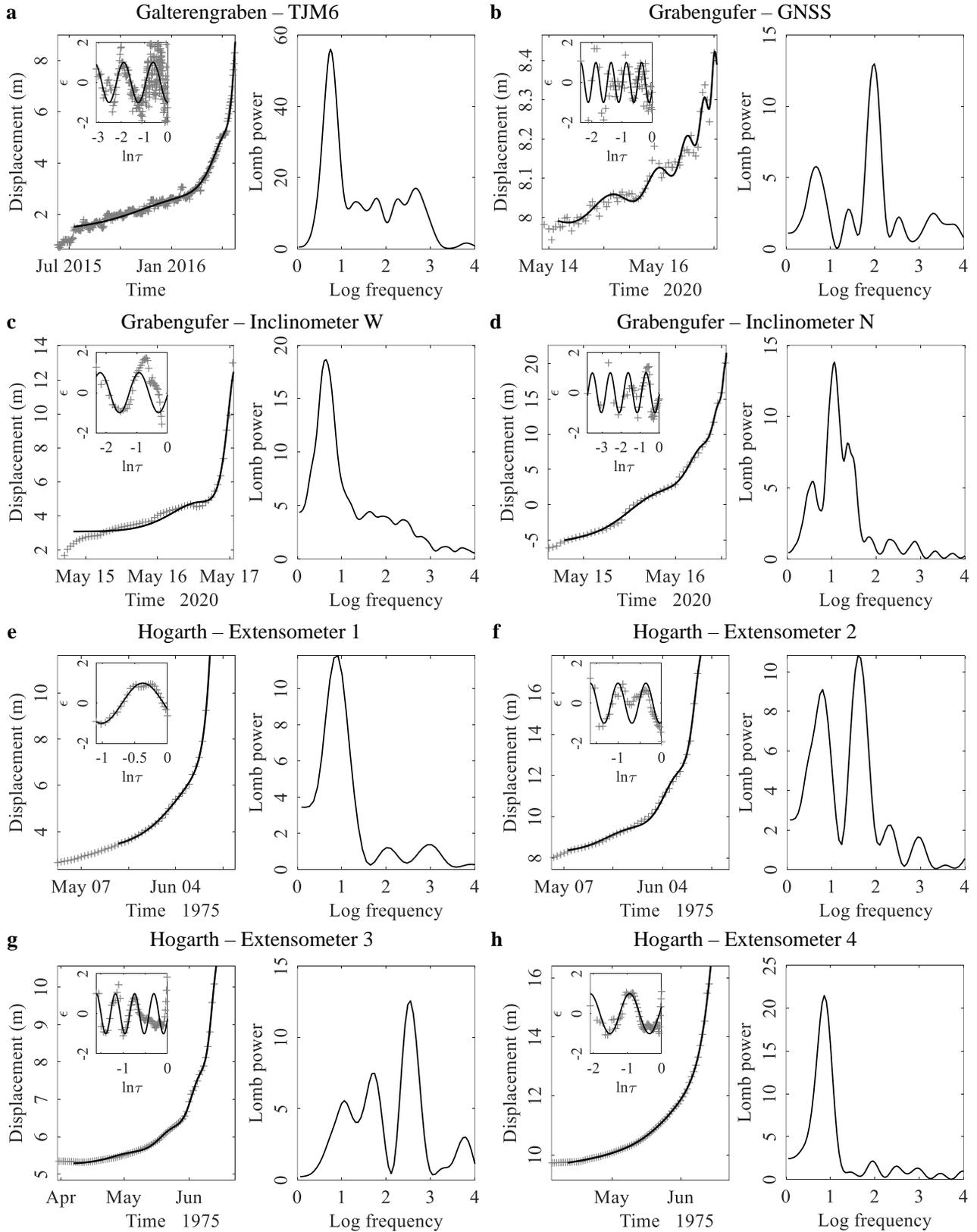

**Figure S6.** Time series of the displacement data of various landslides fitted to the LPPLS model (insets show normalized residual $\epsilon$ as a function of log normalized time $\tau$) and Lomb periodograms during acceleration crises.

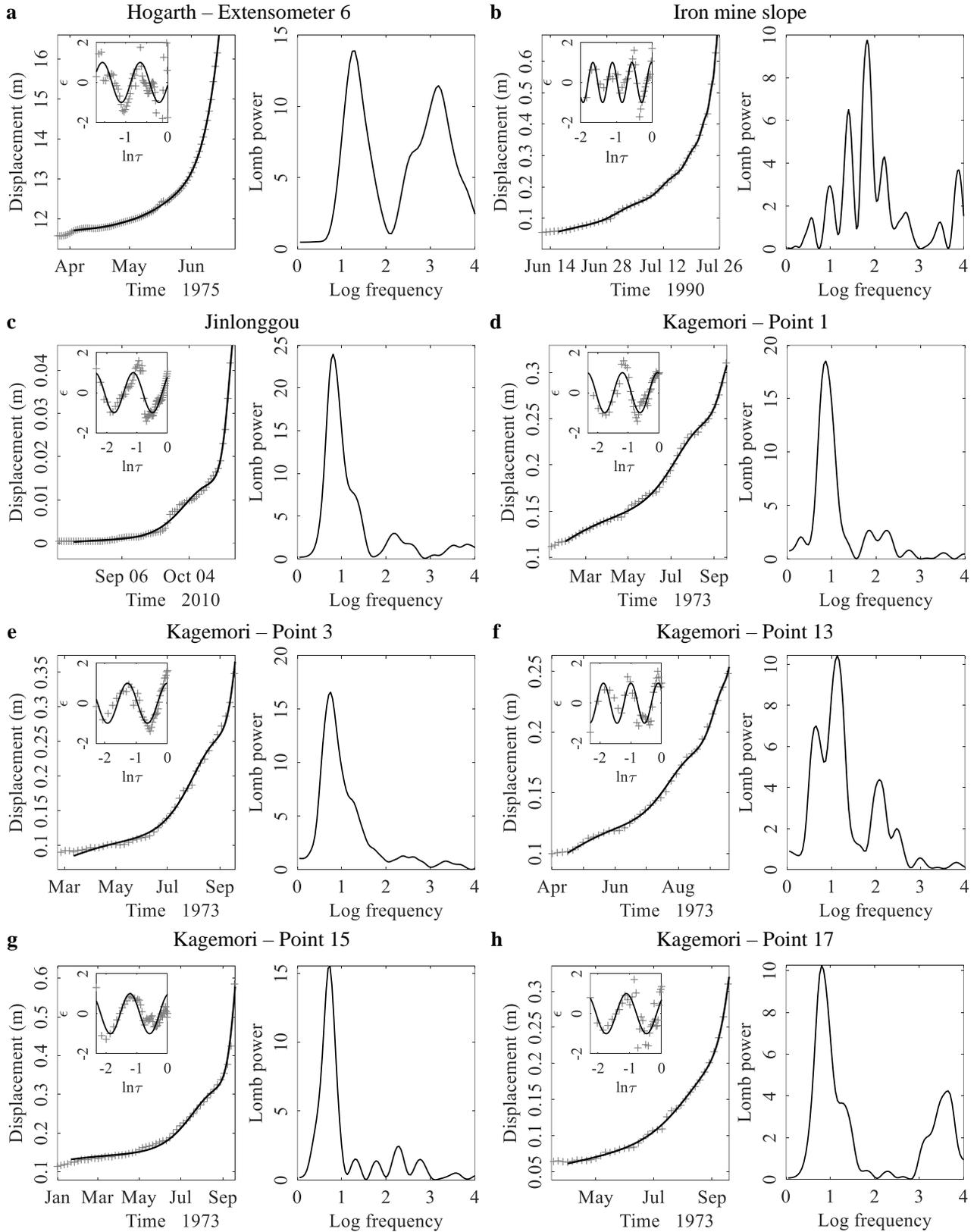

**Figure S7.** Time series of the displacement data of various landslides fitted to the LPPLS model (insets show normalized residual $\epsilon$ as a function of log normalized time $\tau$) and Lomb periodograms during acceleration crises.

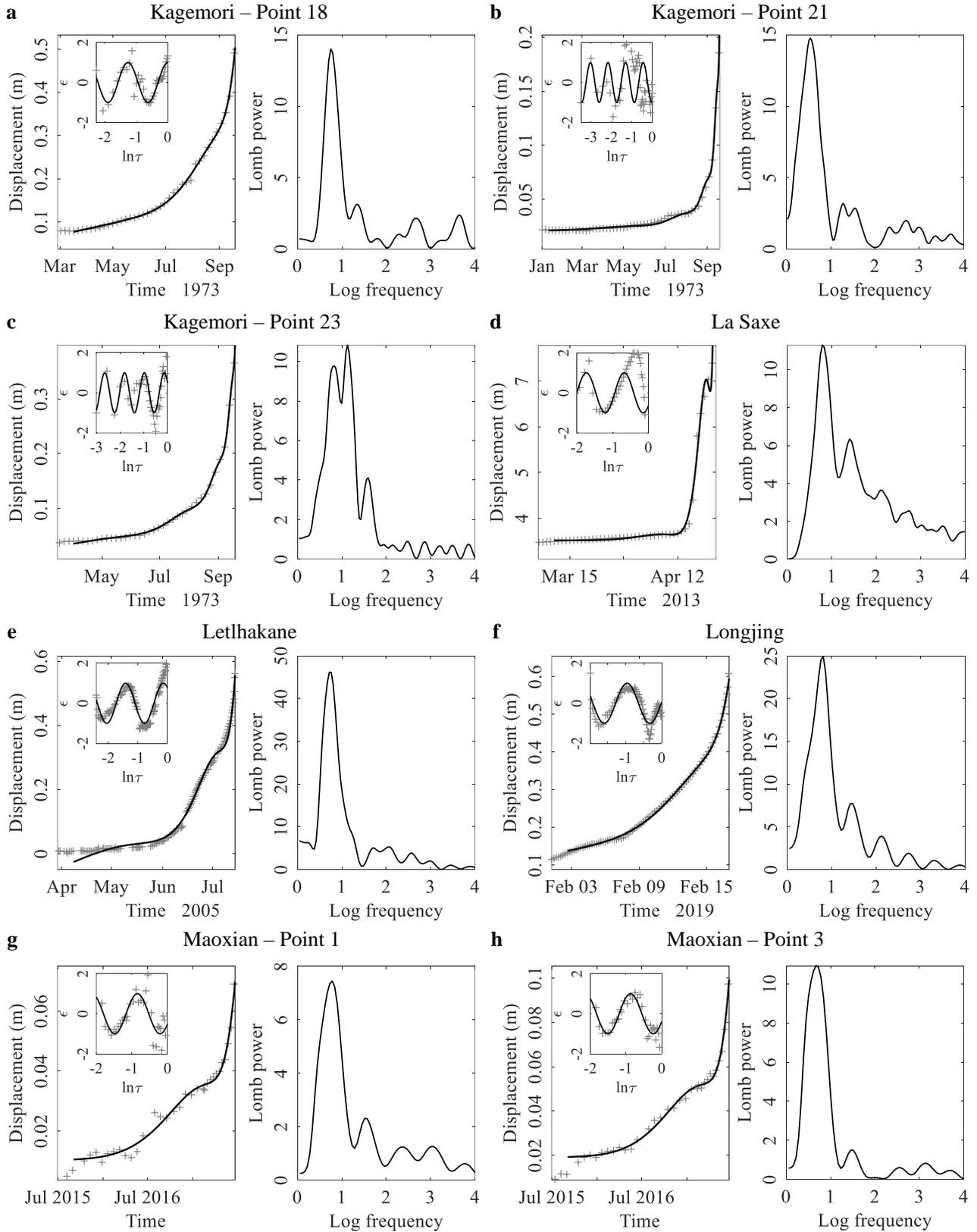

**Figure S8.** Time series of the displacement data of various landslides fitted to the LPPLS model (insets show normalized residual $\epsilon$ as a function of log normalized time $\tau$) and Lomb periodograms during acceleration crises.

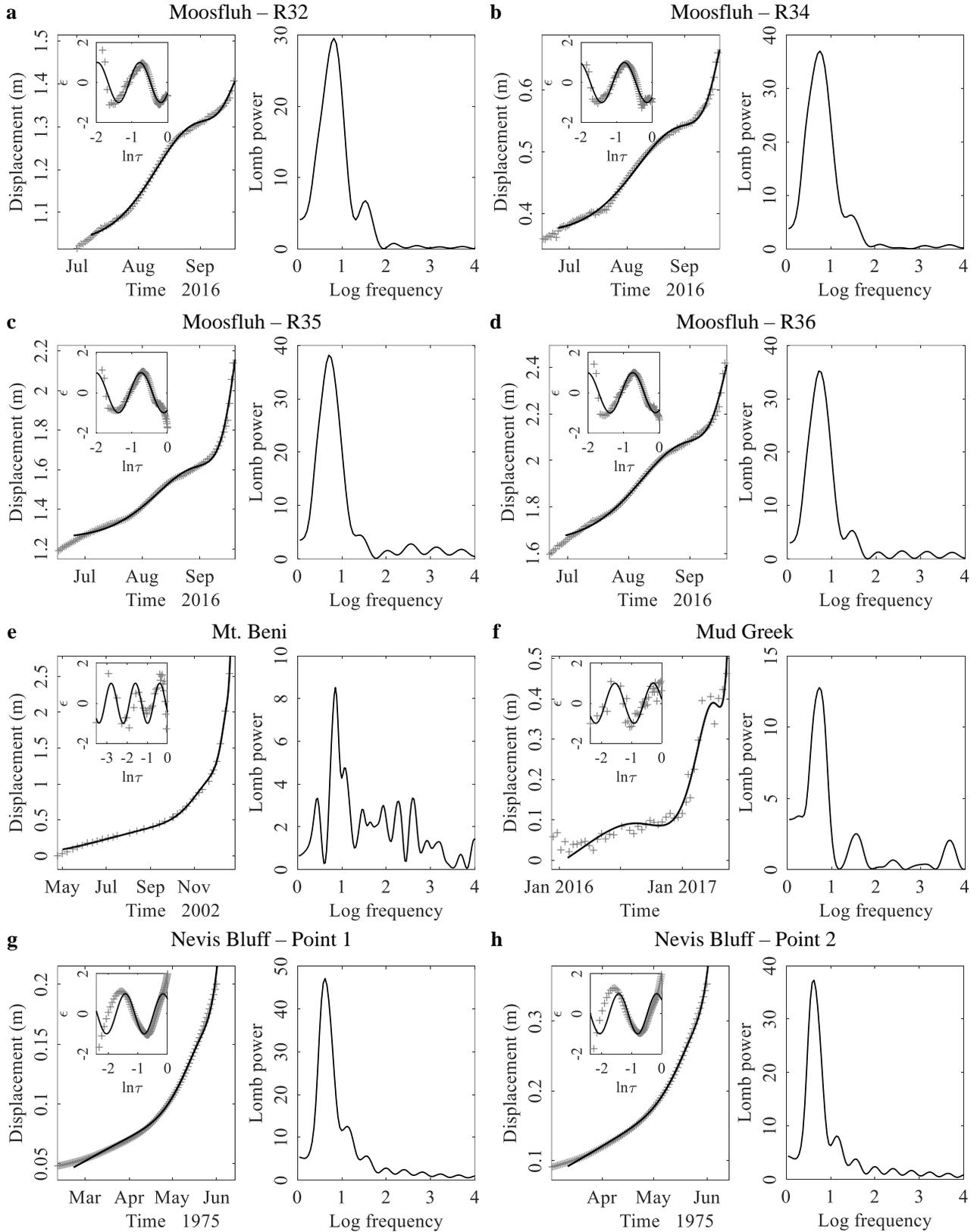

**Figure S9.** Time series of the displacement data of various landslides fitted to the LPPLS model (insets show normalized residual $\epsilon$ as a function of log normalized time $\tau$) and Lomb periodograms during acceleration crises.

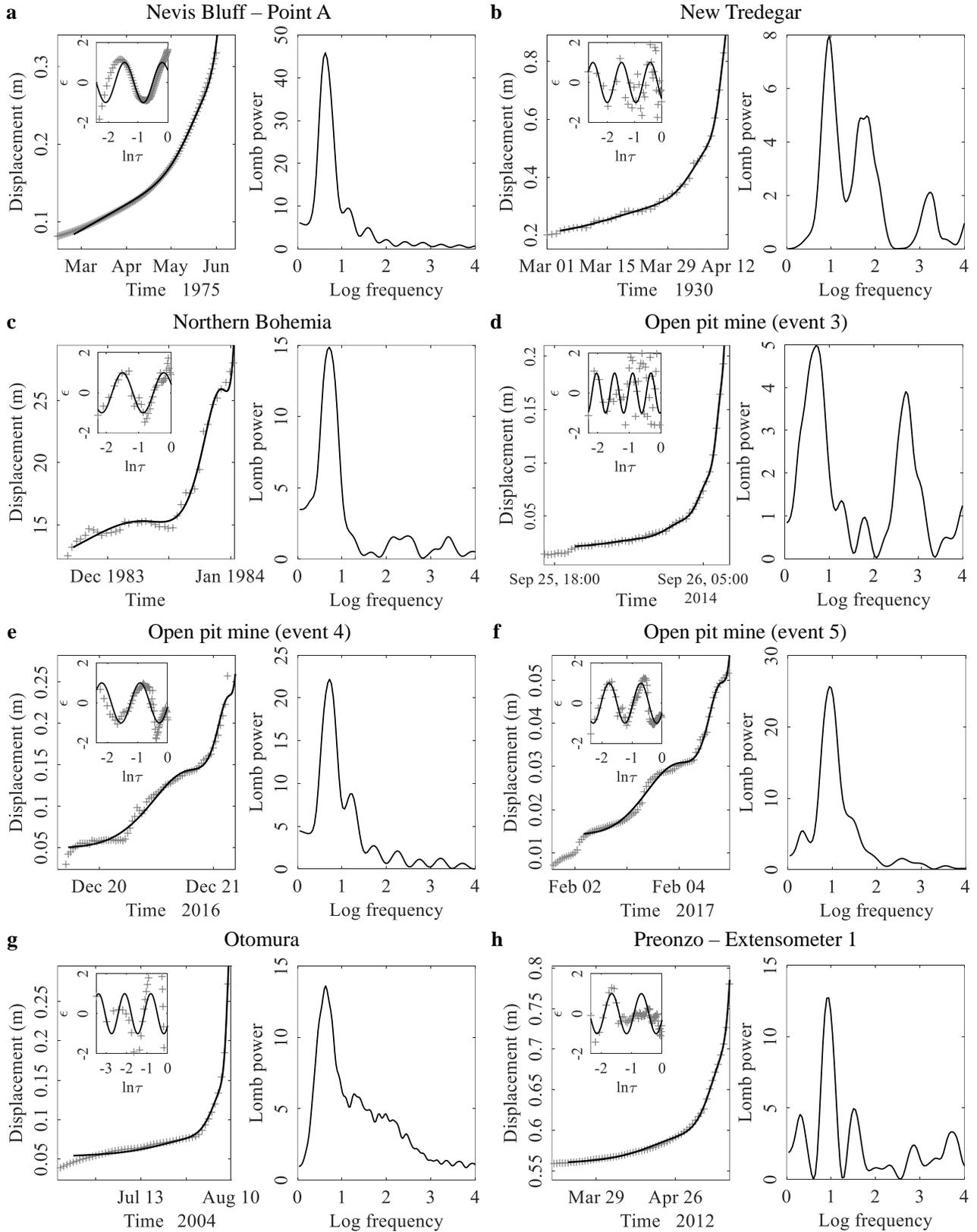

**Figure S10.** Time series of the displacement data of various landslides fitted to the LPPLS model (insets show normalized residual $\epsilon$ as a function of log normalized time $\tau$) and Lomb periodograms during acceleration crises.

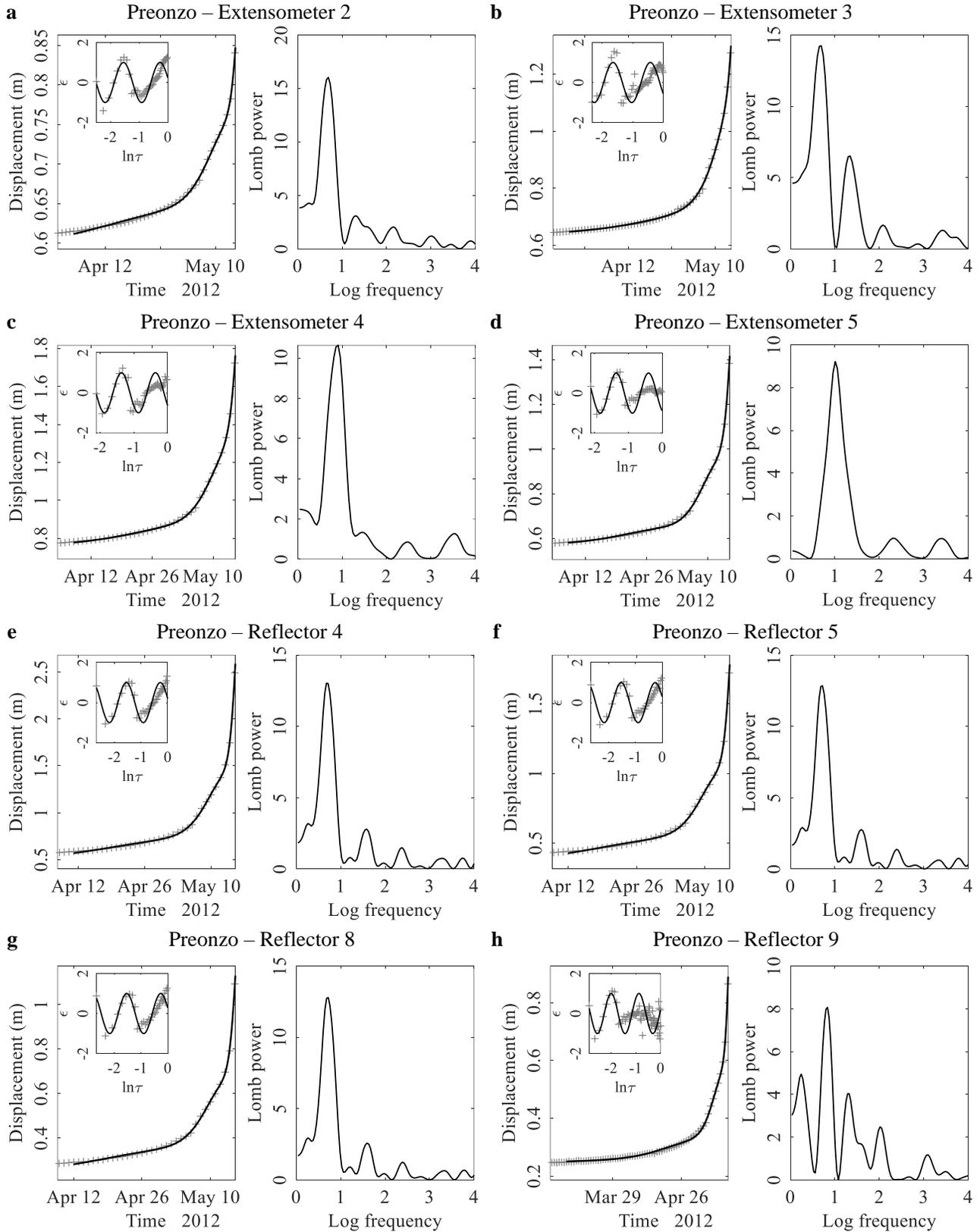

**Figure S11.** Time series of the displacement data of various landslides fitted to the LPPLS model (insets show normalized residual $\epsilon$ as a function of log normalized time $\tau$) and Lomb periodograms during acceleration crises.

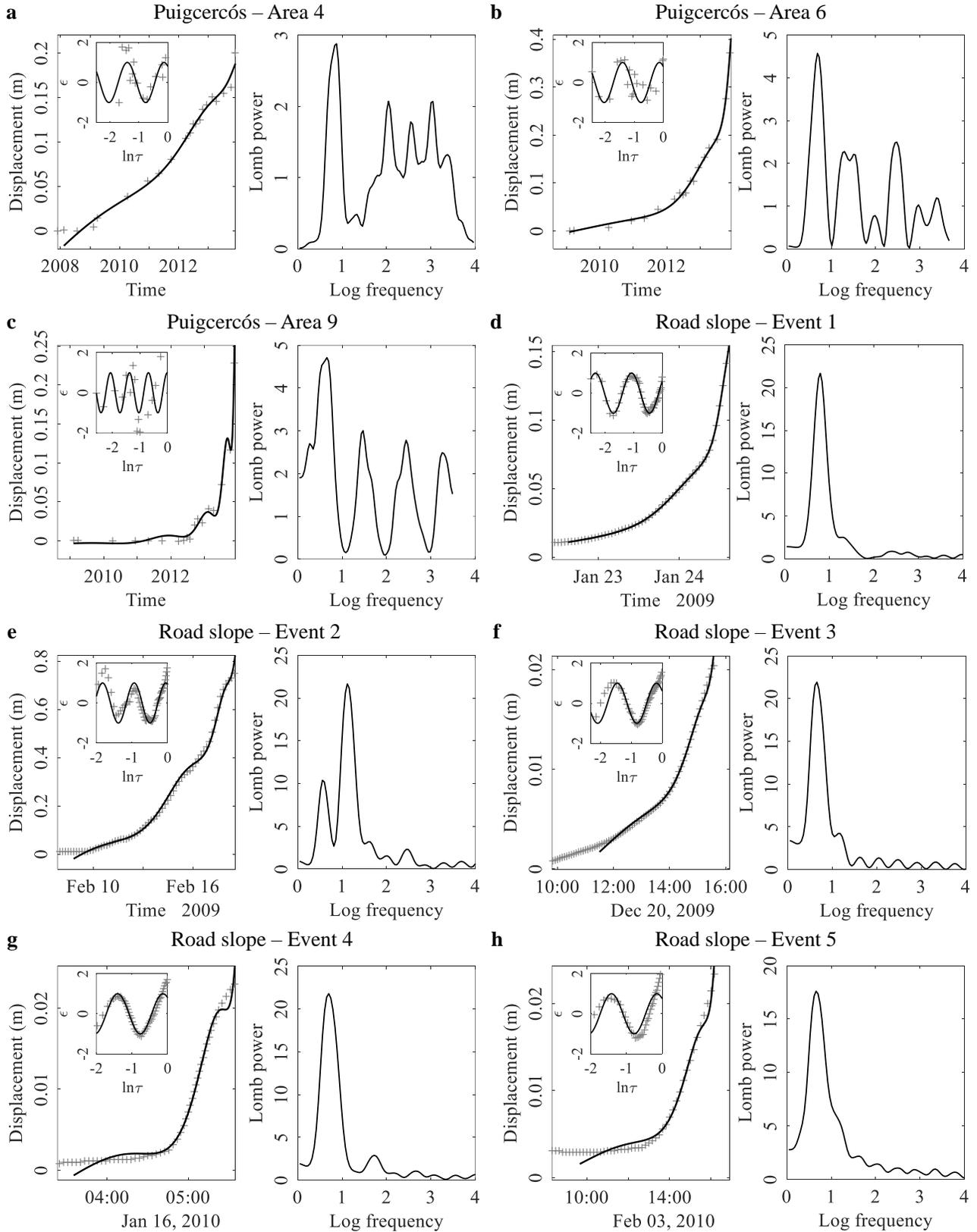

**Figure S12.** Time series of the displacement data of various landslides fitted to the LPPLS model (insets show normalized residual $\epsilon$ as a function of log normalized time $\tau$) and Lomb periodograms during acceleration crises.

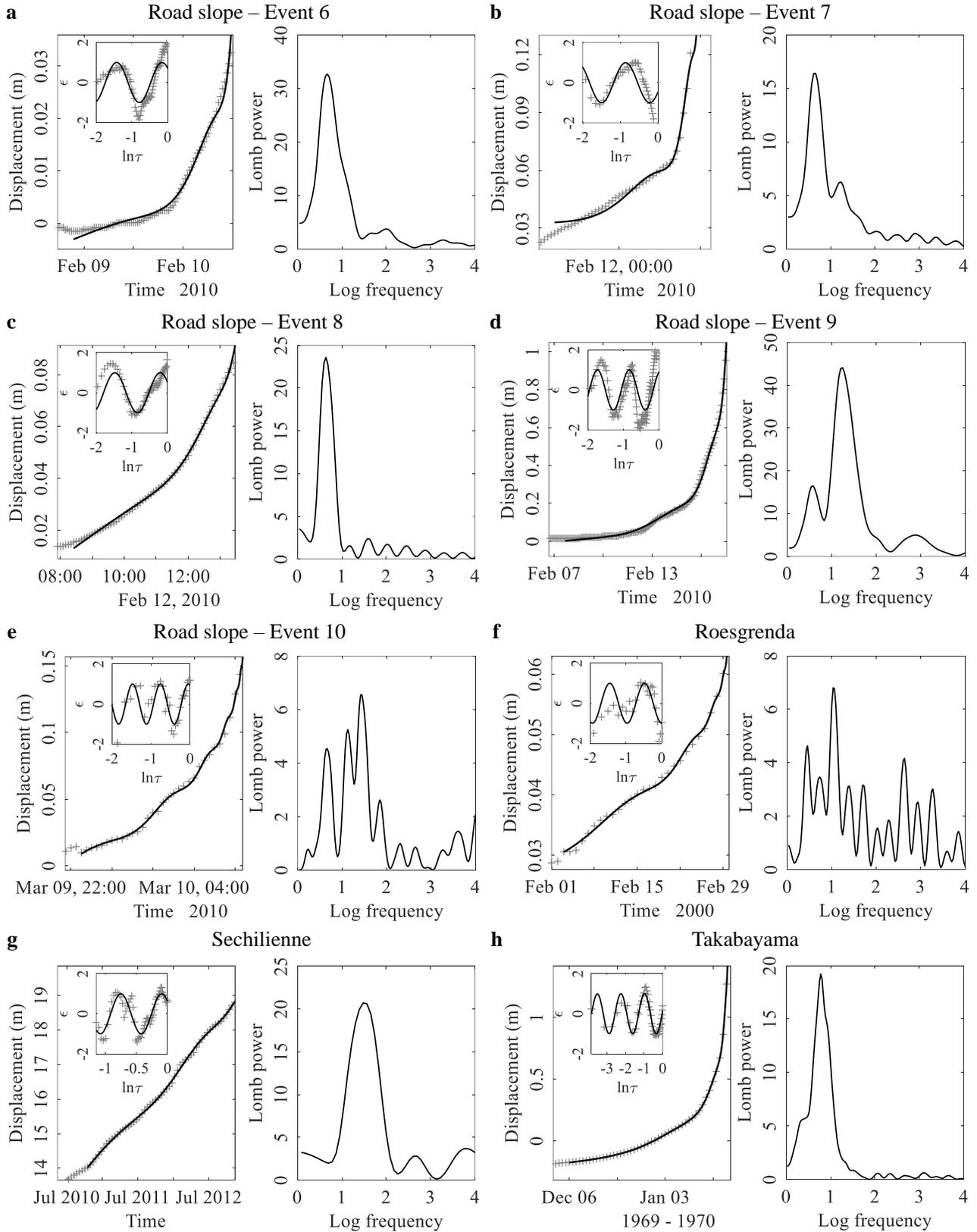

**Figure S13.** Time series of the displacement data of various landslides fitted to the LPPLS model (insets show normalized residual $\epsilon$ as a function of log normalized time $\tau$) and Lomb periodograms during acceleration crises.

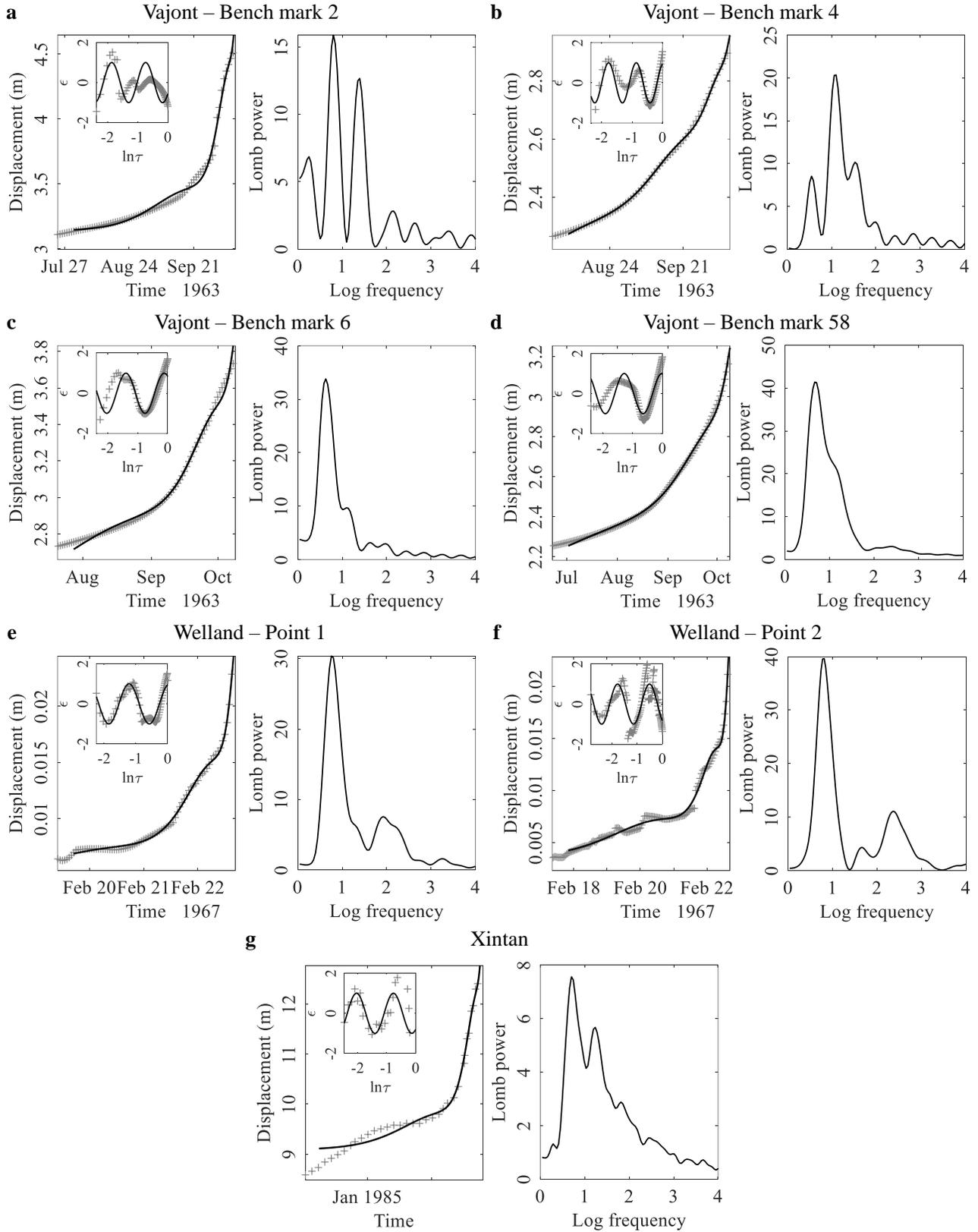

**Figure S14.** Time series of the displacement data of various landslides fitted to the LPPLS model (insets show normalized residual $\epsilon$ as a function of log normalized time $\tau$) and Lomb periodograms during acceleration crises.

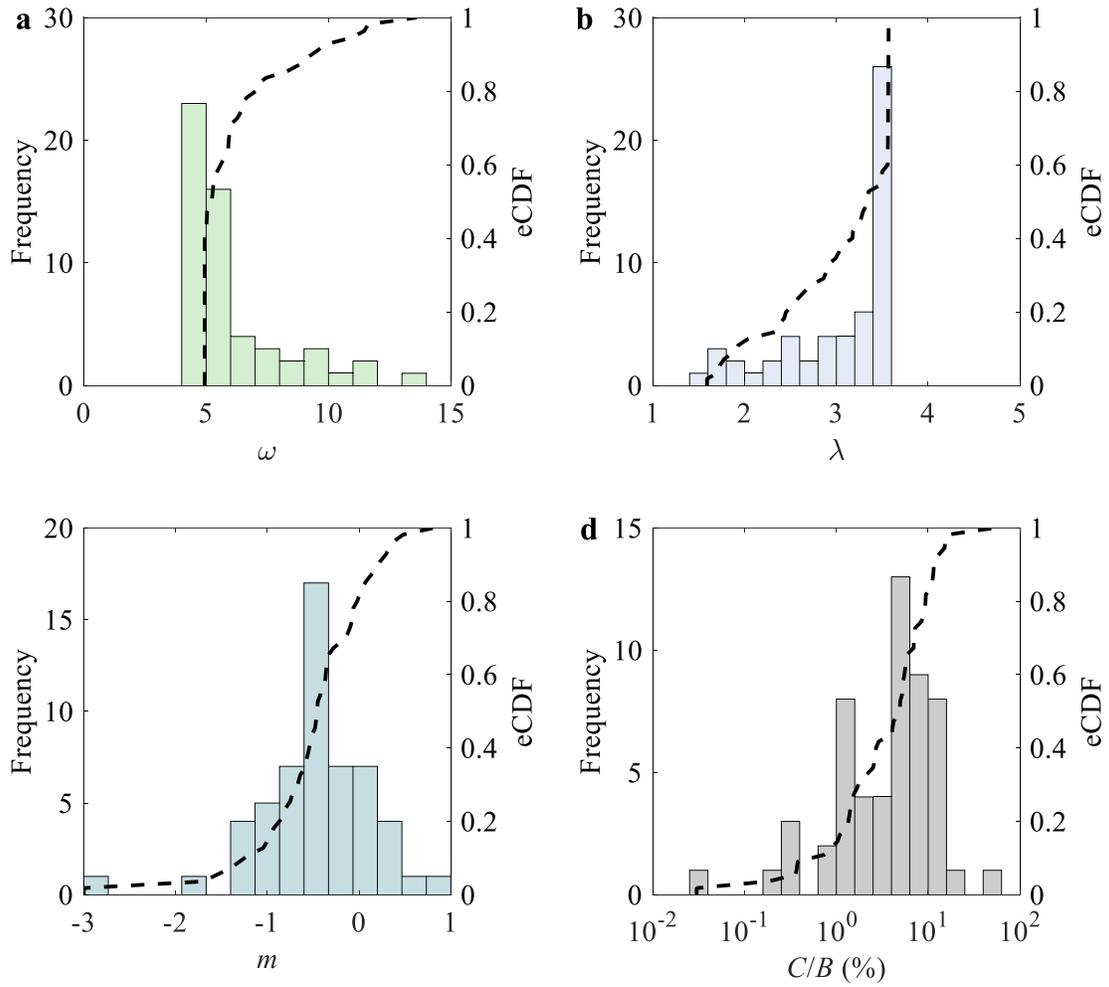

**Figure S15.** Histograms together with the empirical cumulative distribution function (eCDF) of the LPPLS parameters of 52 landslides, including (a) the log-periodic frequency $\omega$, (b) the scaling ratio $\lambda$, (c) the exponent $m$, and (d) the relative amplitude $C/B$.

**Table S1.** Landslide information (52 sites in total).

| Site | Location | Type | Material | Failure time | Volume (m$^3$) | Monitoring method | Data source | Reference |
|---|---|---|---|---|---|---|---|---|
| Abbotsford | New Zealand | Soilslide | Clay | 1979-08-08 | $5×10^6$ | Survey lines | Digitized | (Hancox, 2008) |
| Achoma | Peru | Soilslide | Lacustrine sediments | 2020-06-18 | $5.4×10^6$ | Optical satellites | Digitized | (Lacroix et al., 2023) |
| Agoyama | Japan | Rockslide | Tuffaceous sandstone | 1972-12-02 | $\sim 10^5$ | Geodetic bench marks | Digitized | (Hayashi & Yamamori, 1991) |
| Arvigo | Switzerland | Topple / rockslide | Gneiss | 2007-05-28 | $2×10^5$ | Telejointmeter | Original | (Leinauer et al., 2023) |
| Baishi | China | Rockslide | Phyllite | 2007-07-28 | $2×10^6$ | Total station with reflectors | Digitized | (Tang et al., 2024) |
| Baiyan | China | Rockslide | Limestone | 2022-05-08 | $2.5×10^4$ | Satellite-based InSAR | Digitized | (Li et al., 2023) |
| Brienz / Brinzauls | Switzerland | Rockslide | Flysch, schists, dolomite | 2023-06-15 | $1.2×10^6$ | Total station with reflectors | Original | (Loew et al., 2024) |
| Cadia | Australia | Soilslide | Earthfill materials | 2018-03-09 | $7.2×10^4$ | Satellite-based InSAR | Digitized | (Carlà, Intrieri, et al., 2019) |
| Copper open pit | Undisclosed | Rockslide | Limestone, spilite | 2016-11-17 | $6.4×10^5$ | Satellite-based InSAR | Digitized | (Carlà, Intrieri, et al., 2019) |
| Dosan | Japan | Rockslide | Schist | 1962-02-20 | $6×10^4$ | Crack meter | Digitized | (Saito, 1965) |
| Gallivaggio | Italy | Rockfall | Granite | 2018-05-29 | $5×10^3$ | Ground-based InSAR | Digitized | (Carlà, Nolesini, et al., 2019) |
| Galterengraben | Switzerland | Rockfall | Sandstone | 2016-04-24 | $2.5×10^3$ | Telejointmeter | Original | (Leinauer et al., 2023) |
| Grabengufer | Switzerland | Rockfall | Rock & ice | 2020-05-17 | $5×10^2$ | GNSS & inclinometer | Original | (Cicoira et al., 2022) |
| Hogarth | Canada | Topple | Diorite | 1975-06-23 | $2×10^5$ | Extensometers | Digitized | (Brawner & Stacey, 1979) |
| Iron mine | Mexico | Rockslide | Rock | 1990-07-26 | $\sim 5×10^4$ | Total station with reflectors | Digitized | (Ryan & Call, 1992) |
| Jinlonggou | China | Rockslide | Syenite & basalt | 2010-10-23 | $2×10^5$ | Extensometer | Digitized | (Chen et al., 2021) |
| Kagemori | Japan | Rockslide | Limestone | 1973-09-20 | $3×10^5$-$4×10^5$ | Measuring tapes | Digitized | (Yamaguchi & Shimotani, 1986) |
| La Saxe | Italy | Rockslide | Meta-sedimentary sequences | 2013-04-21 | $5×10^2$-$1×10^3$ | Total station with reflectors | Original | (Manconi & Giordan, 2016) |
| La Clapière | France | Rockslide | Metamorphic rocks | N/A | $5×10^7$ | Distance meters | Original | (Helmstetter et al., 2004) |
| Letlhakane diamond mine | Botswana | Rockslide | Sandstone | 2005-07-14 | $2.3×10^5$ | Total station with reflectors | Digitized | (Kayesa, 2006) |
| Longjing | China | Rockslide | Dolomite & limestone | 2019-02-17 | $1.4×10^6$ | Extensometers | Digitized | (Fan et al., 2019) |
| Maoxian | China | Rockslide | Sandstone, phyllite | 2017-06-24 | $1.5×10^7$ | Satellite-based InSAR | Digitized | (Intrieri et al., 2018) |
| Moosfluh | Switzerland | Rockslide | Metamorphic rocks | N/A | $7.5×10^7$ | Total station with reflectors | Original | (Glueer et al., 2019) |

**Table S1 (continued).** Landslide information (52 sites in total).

| Site | Location | Type | Material | Failure time | Volume (m$^3$) | Monitoring method | Data source | Reference |
|---|---|---|---|---|---|---|---|---|
| Mt. Beni | Italy | Rockslide | Basalt & limestone | 2002-12-28 | $5.0 \times 10^5$ | Distometric benchmarks | Digitized | (Gigli et al., 2011) |
| Mud Greek | USA | Rockslide | Shale, sandstone, sediments | 2017-05-20 | $3 \times 10^6$ | Satellite-based InSAR | Digitized | (Jacquemart & Tiampo, 2021) |
| Nevis Bluff | New Zealand | Flexural topple / rockslide | Schist | 1975-06-14 | $3.2 \times 10^4$ | Survey markers | Digitized | (Brown et al., 1980) |
| New Tredegar | UK | Rockslide | Sandstone | 1930-04-12 | $\sim 7 \times 10^4$ | Unspecified | Digitized | (Carey, 2011) |
| Northern Bohemia | Czech Republic | Rockfall | Sandstone | 1984-01-07 | $1.4 \times 10^3$ | Extensometers | Digitized | (Zvelebill & Moser, 2001) |
| Open pit mine (event 3) | Undisclosed | Rockslide | Anorthosite | 2014-09-26 | $6 \times 10^2$ | Ground-based InSAR | Digitized | (Carlà et al., 2017) |
| Open pit mine (event 4) | Undisclosed | Rockslide | Anorthosite | 2014-2017 | $3 \times 10^3$ | Ground-based InSAR | Digitized | (Carlà et al., 2017) |
| Open pit mine (event 5) | Undisclosed | Topple | Anorthosite | 2017-02-05 | $4 \times 10^3$ | Ground-based InSAR | Digitized | (Carlà et al., 2017) |
| Otomura | Japan | Rockslide | Sandstone & shale | 2004-08-10 | $2 \times 10^5$ | Extensometers | Digitized | (Fujisawa et al., 2010) |
| Preonzo | Switzerland | Rockslide | Gneiss | 2012-05-15 | $2.1 \times 10^5$ | Extensometers & total station with reflectors | Original | (Loew et al., 2017) |
| Puigcercós | Spain | Rockfall | Marl, silt, sandstone, limestone | 2013-12-03 | $1 \times 10^3$ | LiDAR | Digitized | (Royán et al., 2015) |
| Road slope (event 1) | Undisclosed | Rockslide | Mobilized gneiss | 2009-01-24 | $3 \times 10^2$ | Ground-based InSAR | Digitized | (Mazzanti et al., 2015) |
| Road slope (event 2) | Undisclosed | Flow | Colluvium | 2009-02-18 | $1.4 \times 10^1$ | Ground-based InSAR | Digitized | (Mazzanti et al., 2015) |
| Road slope (event 3) | Undisclosed | Soilslide | Colluvium & beton | 2009-12-20 | $1.6 \times 10^2$ | Ground-based InSAR | Digitized | (Mazzanti et al., 2015) |
| Road slope (event 4) | Undisclosed | Soilslide | Colluvium & beton | 2010-01-16 | $2 \times 10^2$ | Ground-based InSAR | Digitized | (Mazzanti et al., 2015) |
| Road slope (event 5) | Undisclosed | Soilslide | Colluvium & beton | 2009-02-03 | $8 \times 10^1$ | Ground-based InSAR | Digitized | (Mazzanti et al., 2015) |
| Road slope (event 6) | Undisclosed | Soilslide | Colluvium & beton | 2010-02-11 | $5 \times 10^2$ | Ground-based InSAR | Digitized | (Mazzanti et al., 2015) |
| Road slope (event 7) | Undisclosed | Flow | Mobilized & altered gneiss | 2010-02-12 | $2 \times 10^2$ | Ground-based InSAR | Digitized | (Mazzanti et al., 2015) |
| Road slope (event 8) | Undisclosed | Rockslide | Colluvium & beton | 2010-02-12 | $3 \times 10^2$ | Ground-based InSAR | Digitized | (Mazzanti et al., 2015) |
| Road slope (event 9) | Undisclosed | Rockslide | Mobilized & altered gneiss | 2010-02-17 | $8 \times 10^1$ | Ground-based InSAR | Digitized | (Mazzanti et al., 2015) |
| Road slope (event 10) | Undisclosed | Rockslide | Mobilized & altered gneiss | 2010-03-10 | $1.5 \times 10^2$ | Ground-based InSAR | Digitized | (Mazzanti et al., 2015) |

**Table S1 (continued).** Landslide information (52 sites in total).

| Site | Location | Type | Material | Failure time | Volume (m$^3$) | Monitoring method | Data source | Reference |
| --- | --- | --- | --- | --- | --- | --- | --- | --- |
| Roesgrenda | Norway | Soilslide | Quick clay | 2000-03-02 | $2\times10^3$ | Extensometers | Digitized | (Okamoto et al., 2004) |
| Ruinon | Italy | Rockslide | Phyllite | N/A | $1.3\times10^7$ | Distometers & extensometers | Digitized | (Crosta & Agliardi, 2002, 2003) |
| Séchilienne | France | Rockslide | Micaschists | N/A | $\sim4\times10^6$ | Extensometer | Original | (Helmstetter & Garambois, 2010) |
| Takabayama | Japan | Rockslide | Mudstone, sandstone | 1970-01-22 | $5\times10^3$ | Extensometers | Digitized | (Saito, 1979) |
| Vajont | Italy | Rockslide | Limestone | 1963-10-09 | $2.7\times10^8$ | Geodetic bench marks | Digitized | (Nonveiller, 1987) |
| Veslemannen | Norway | Rockslide | Gneiss | 2019-09-05 | $5.4\times10^4$ | Ground-based InSAR | Original | (Kristensen et al., 2021) |
| Welland | Canada | Soilslide | Clay | 1967-02-22 | $5\times10^2$ | Extensometers | Digitized | (Kwan, 1971) |
| Xintan | China | Rockslide | Sediments | 1985-06-12 | $3\times10^7$ | Geodetic bench marks | Digitized | (Xue et al., 2014) |

Note: InSAR - Interferometric Synthetic Aperture Radar; LiDAR - Light Detection and Ranging; GNSS - Global navigation satellite system.

**Table S2.** Parameters of the LPPLS calibration and Lomb analysis for the different landslides.

| Site | $m_{LPPLS}$ | $\omega_{LPPLS}$ | $\lambda_{LPPLS}$ | $C/B$ (%) | $f_{Lomb}$ | $\omega_{Lomb}$ | $\lambda_{Lomb}$ | $P_{max}$ | $\eta$ | $p_{FA}$ | $\gamma$ |
|---|---|---|---|---|---|---|---|---|---|---|---|
| Abbotsford | -0.73 | 4.94 | 3.57 | 4.69 | 0.72 | 4.55 | 3.98 | 16.10 | 3.67 | 0.00 | 3.67 |
| Achoma | 0.076 | 5.01 | 3.50 | 1.03 | 0.79 | 4.99 | 3.52 | 5.80 | 1.38 | 0.10 | 1.02 |
| Agoyama | -0.35 | 5.27 | 3.29 | 1.09 | 0.85 | 5.35 | 3.24 | 10.81 | 1.31 | 0.00 | 1.42 |
| Arvigo | -0.063 | 5.04 | 3.48 | 0.38 | 0.82 | 5.16 | 3.38 | 11.85 | 1.49 | 0.00 | 1.10 |
| Baishi | 0.0036 | 4.94 | 3.57 | 0.03 | 0.71 | 4.46 | 4.09 | 33.44 | 4.78 | 0.00 | 3.47 |
| Baiyan | 0.33 | 4.94 | 3.57 | 9.23 | 0.65 | 4.10 | 4.63 | 12.52 | 8.98 | 0.00 | 3.18 |
| Brienz/Brinzauls (715) | -0.53 | 4.94 | 3.57 | 1.22 | 0.68 | 4.24 | 4.40 | 52.04 | 4.20 | 0.00 | 2.41 |
| Brienz/Brinzauls (719) | -0.63 | 4.94 | 3.57 | 1.44 | 0.64 | 4.04 | 4.73 | 55.14 | 7.13 | 0.00 | 2.27 |
| Brienz/Brinzauls (725) | -0.61 | 4.94 | 3.57 | 1.40 | 0.67 | 4.23 | 4.41 | 51.16 | 4.02 | 0.00 | 2.19 |
| Cadia | -0.8 | 4.94 | 3.57 | 5.64 | 0.66 | 4.13 | 4.58 | 6.55 | 1.90 | 0.03 | 1.63 |
| Copper open pit | -1.2 | 9.88 | 1.89 | 16.35 | 0.84 | 5.25 | 3.31 | 1.94 | 1.15 | 0.98 | 0.62 |
| Dosan (point 1) | -0.15 | 9.41 | 1.95 | 0.98 | 1.49 | 9.36 | 1.96 | 12.55 | 1.48 | 0.00 | 1.34 |
| Dosan (point 2) | -0.52 | 4.94 | 3.57 | 2.37 | 0.70 | 4.39 | 4.18 | 14.58 | 7.90 | 0.00 | 2.05 |
| Dosan (point 3) | -0.38 | 4.94 | 3.57 | 0.83 | 1.80 | 11.31 | 1.74 | 10.03 | 1.06 | 0.00 | 1.22 |
| Gallivaggio | -1.2 | 11.63 | 1.72 | 7.20 | 1.40 | 8.77 | 2.05 | 5.57 | 1.04 | 0.16 | 0.81 |
| Galterengraben (TJM1) | -0.49 | 5.99 | 2.85 | 3.93 | 0.93 | 5.83 | 2.94 | 123.79 | 3.78 | 0.00 | 2.09 |
| Galterengraben (TJM2) | -0.53 | 4.94 | 3.57 | 4.24 | 0.77 | 4.82 | 3.68 | 99.28 | 2.72 | 0.00 | 1.64 |
| Galterengraben (TJM6) | -0.51 | 4.94 | 3.57 | 4.42 | 0.75 | 4.68 | 3.82 | 55.99 | 3.30 | 0.00 | 1.12 |
| Grabengufer (GNSS) | -0.35 | 12.42 | 1.66 | 6.43 | 1.99 | 12.50 | 1.65 | 12.97 | 2.25 | 0.00 | 1.14 |
| Grabengufer (inclino. W) | -1.2 | 4.94 | 3.57 | 21.25 | 0.65 | 4.07 | 4.67 | 18.65 | 4.21 | 0.00 | 2.18 |
| Grabengufer (inclino. N) | 0.1 | 6.70 | 2.55 | 0.73 | 1.06 | 6.68 | 2.56 | 13.81 | 1.70 | 0.00 | 1.90 |
| Hogarth (extensometer 1) | -2.7 | 4.94 | 3.57 | 25.54 | 0.91 | 5.75 | 2.98 | 11.82 | 8.61 | 0.00 | 3.75 |
| Hogarth (extensometer 2) | -1.3 | 9.86 | 1.89 | 6.66 | 1.59 | 10.01 | 1.87 | 10.81 | 1.19 | 0.00 | 1.62 |
| Hogarth (extensometer 3) | -1.8 | 14.45 | 1.54 | 5.64 | 2.55 | 16.01 | 1.48 | 12.57 | 1.68 | 0.00 | 1.08 |
| Hogarth (extensometer 4) | -0.87 | 5.39 | 3.21 | 2.39 | 0.86 | 5.40 | 3.20 | 21.48 | 10.02 | 0.00 | 1.97 |
| Hogarth (extensometer 6) | -1.6 | 6.98 | 2.46 | 2.21 | 1.29 | 8.14 | 2.16 | 13.89 | 1.21 | 0.00 | 1.11 |
| Iron mine | -0.13 | 11.46 | 1.73 | 0.36 | 1.83 | 11.48 | 1.73 | 9.75 | 1.50 | 0.00 | 1.38 |
| Jinlonggou | -0.87 | 4.94 | 3.57 | 11.51 | 0.81 | 5.06 | 3.46 | 23.94 | 7.98 | 0.00 | 2.30 |
| Kagemori (point 1) | 0.26 | 5.50 | 3.13 | 2.04 | 0.87 | 5.45 | 3.17 | 18.51 | 6.90 | 0.00 | 2.72 |
| Kagemori (point 3) | -0.23 | 4.94 | 3.57 | 2.46 | 0.75 | 4.71 | 3.80 | 16.59 | 13.02 | 0.00 | 3.37 |
| Kagemori (point 13) | 0.064 | 7.07 | 2.43 | 0.33 | 1.12 | 7.07 | 2.43 | 10.42 | 1.49 | 0.00 | 1.93 |
| Kagemori (point 15) | -0.77 | 4.94 | 3.57 | 9.09 | 0.73 | 4.62 | 3.90 | 15.50 | 6.35 | 0.00 | 1.86 |
| Kagemori (point 17) | -0.42 | 4.94 | 3.57 | 1.99 | 0.80 | 5.03 | 3.48 | 10.25 | 2.42 | 0.00 | 1.68 |
| Kagemori (point 18) | -0.52 | 4.94 | 3.57 | 3.57 | 0.75 | 4.72 | 3.79 | 14.02 | 4.50 | 0.00 | 2.37 |
| Kagemori (point 21) | -1.1 | 7.29 | 2.37 | 9.38 | 0.54 | 3.41 | 6.31 | 14.78 | 4.65 | 0.00 | 1.66 |
| Kagemori (point 23) | -0.66 | 7.49 | 2.31 | 3.70 | 1.13 | 7.09 | 2.43 | 10.82 | 1.11 | 0.00 | 1.80 |
| La Clapière | 0.23 | 6.28 | 2.72 | 1.54 | 1.02 | 6.41 | 2.67 | 11.41 | 1.47 | 0.00 | 2.36 |
| La Saxe | -3 | 5.91 | 2.90 | 52.94 | 0.79 | 4.95 | 3.56 | 11.29 | 1.78 | 0.00 | 1.52 |
| Letlhakane diamond mine | -0.37 | 4.94 | 3.57 | 5.76 | 0.72 | 4.52 | 4.01 | 46.17 | 7.33 | 0.00 | 3.82 |
| Longjing | -0.45 | 4.94 | 3.57 | 2.63 | 0.81 | 5.09 | 3.44 | 24.91 | 3.22 | 0.00 | 1.99 |
| Maoxian (point 1) | -0.82 | 4.94 | 3.57 | 12.93 | 0.78 | 4.93 | 3.58 | 7.45 | 3.23 | 0.02 | 1.36 |
| Maoxian (point 2) | -1 | 4.94 | 3.57 | 16.48 | 0.67 | 4.23 | 4.42 | 10.76 | 6.57 | 0.00 | 2.13 |
| Maoxian (point 3) | -0.98 | 4.94 | 3.57 | 16.55 | 0.67 | 4.23 | 4.42 | 10.94 | 7.27 | 0.00 | 2.19 |
| Moosfluh (reflector 32) | 0.73 | 5.29 | 3.28 | 10.00 | 0.82 | 5.16 | 3.38 | 29.52 | 4.38 | 0.00 | 2.81 |

**Table S2 (continued).** Parameters of the LPPLS calibration and Lomb analysis for the different landslides.

| Site | $m_{LPPLS}$ | $\omega_{LPPLS}$ | $\lambda_{LPPLS}$ | $C/B$ (%) | $f_{Lomb}$ | $\omega_{Lomb}$ | $\lambda_{Lomb}$ | $P_{max}$ | $\eta$ | $p_{FA}$ | $\gamma$ |
|---|---|---|---|---|---|---|---|---|---|---|---|
| Moosfluh (reflector 34) | 0.022 | 4.94 | 3.57 | 0.35 | 0.76 | 4.79 | 3.71 | 36.99 | 5.85 | 0.00 | 3.37 |
| Moosfluh (reflector 35) | -0.61 | 4.94 | 3.57 | 9.29 | 0.71 | 4.45 | 4.11 | 38.14 | 8.47 | 0.00 | 3.77 |
| Moosfluh (reflector 36) | -0.003 | 4.94 | 3.57 | 0.05 | 0.71 | 4.45 | 4.10 | 35.18 | 6.57 | 0.00 | 3.48 |
| Mt. Beni | -0.33 | 5.21 | 3.34 | 1.78 | 0.85 | 5.37 | 3.22 | 8.51 | 1.79 | 0.01 | 1.51 |
| Mud Greek | -0.46 | 4.94 | 3.57 | 11.12 | 0.72 | 4.53 | 4.00 | 12.77 | 3.39 | 0.00 | 2.02 |
| Nevis Bluff (point 1) | -0.11 | 4.94 | 3.57 | 0.59 | 0.62 | 3.91 | 4.98 | 47.14 | 3.75 | 0.00 | 5.30 |
| Nevis Bluff (point 2) | -0.051 | 4.94 | 3.57 | 0.20 | 0.63 | 3.95 | 4.91 | 37.27 | 4.59 | 0.00 | 4.47 |
| Nevis Bluff (point A) | -0.077 | 4.94 | 3.57 | 0.34 | 0.62 | 3.92 | 4.97 | 45.84 | 4.82 | 0.00 | 5.39 |
| New Tredegar | -0.58 | 5.93 | 2.89 | 2.61 | 0.96 | 6.02 | 2.84 | 7.96 | 1.60 | 0.01 | 1.17 |
| Northern Bohemia | -0.68 | 4.94 | 3.57 | 15.13 | 0.71 | 4.48 | 4.06 | 14.86 | 9.19 | 0.00 | 3.08 |
| Open pit mine (event 3) | -1.4 | 10.94 | 1.78 | 4.09 | 0.70 | 4.41 | 4.15 | 4.98 | 1.28 | 0.26 | 0.76 |
| Open pit mine (event 4) | -0.35 | 5.30 | 3.27 | 5.31 | 0.88 | 5.52 | 3.12 | 23.51 | 4.31 | 0.00 | 2.12 |
| Open pit mine (event 5) | -0.1 | 5.97 | 2.86 | 1.52 | 0.94 | 5.88 | 2.91 | 25.66 | 4.73 | 0.00 | 2.56 |
| Otomura | -0.81 | 4.94 | 3.57 | 5.35 | 0.63 | 3.98 | 4.85 | 13.61 | 2.25 | 0.00 | 1.58 |
| Preonzo (extensometer 1) | -1.1 | 6.47 | 2.64 | 6.03 | 0.95 | 5.94 | 2.88 | 12.73 | 2.57 | 0.00 | 1.27 |
| Preonzo (extensometer 2) | -0.65 | 4.94 | 3.57 | 4.82 | 0.69 | 4.34 | 4.25 | 16.03 | 3.76 | 0.00 | 2.68 |
| Preonzo (extensometer 3) | -1.4 | 5.16 | 3.38 | 6.16 | 0.69 | 4.35 | 4.24 | 14.24 | 2.19 | 0.00 | 1.56 |
| Preonzo (extensometer 4) | -1.5 | 6.63 | 2.58 | 6.59 | 0.95 | 5.99 | 2.85 | 16.65 | 3.92 | 0.00 | 1.94 |
| Preonzo (extensometer 5) | -1.3 | 5.97 | 2.86 | 8.99 | 0.81 | 5.10 | 3.43 | 14.88 | 1.97 | 0.00 | 1.30 |
| Preonzo (reflector 2) | -0.93 | 4.94 | 3.57 | 8.71 | 0.67 | 4.23 | 4.41 | 12.60 | 4.68 | 0.00 | 2.16 |
| Preonzo (reflector 4) | -0.87 | 4.94 | 3.57 | 7.45 | 0.68 | 4.27 | 4.36 | 13.03 | 4.09 | 0.00 | 2.56 |
| Preonzo (reflector 5) | -0.79 | 4.94 | 3.57 | 6.96 | 0.72 | 4.53 | 4.01 | 12.85 | 4.42 | 0.00 | 2.65 |
| Preonzo (reflector 8) | -0.76 | 4.94 | 3.57 | 6.55 | 0.72 | 4.53 | 4.01 | 12.79 | 4.34 | 0.00 | 2.63 |
| Preonzo (reflector 9) | -1.1 | 5.61 | 3.06 | 7.54 | 0.83 | 5.24 | 3.32 | 8.07 | 1.63 | 0.02 | 0.80 |
| Puigcercós (area 4) | 0.63 | 4.94 | 3.57 | 4.36 | 0.88 | 5.50 | 3.13 | 2.88 | 1.39 | 0.67 | 0.93 |
| Puigcercós (area 6) | -0.67 | 4.94 | 3.57 | 5.90 | 0.69 | 4.34 | 4.26 | 4.57 | 1.83 | 0.17 | 1.44 |
| Puigcercós (area 7) | -0.5 | 4.94 | 3.57 | 8.35 | 0.73 | 4.58 | 3.94 | 7.43 | 3.40 | 0.01 | 2.40 |
| Puigcercós (area 9) | -1.7 | 9.16 | 1.99 | 27.41 | 0.66 | 4.14 | 4.57 | 4.71 | 1.57 | 0.15 | 1.48 |
| Road slope (event 1) | -0.41 | 4.94 | 3.57 | 2.70 | 0.80 | 5.00 | 3.52 | 21.67 | 8.66 | 0.00 | 4.31 |
| Road slope (event 2) | -0.021 | 7.08 | 2.43 | 0.18 | 1.12 | 7.02 | 2.45 | 21.65 | 2.09 | 0.00 | 2.43 |
| Road slope (event 3) | 0.0043 | 4.94 | 3.57 | 0.03 | 0.65 | 4.10 | 4.63 | 21.91 | 5.18 | 0.00 | 4.12 |
| Road slope (event 4) | -0.51 | 4.94 | 3.57 | 10.52 | 0.70 | 4.40 | 4.17 | 21.80 | 7.53 | 0.00 | 4.45 |
| Road slope (event 5) | -0.51 | 4.94 | 3.57 | 7.22 | 0.66 | 4.14 | 4.56 | 17.61 | 8.09 | 0.00 | 3.84 |
| Road slope (event 6) | -0.46 | 4.94 | 3.57 | 4.96 | 0.68 | 4.25 | 4.39 | 32.70 | 8.84 | 0.00 | 3.51 |
| Road slope (event 7) | -0.64 | 4.94 | 3.57 | 9.45 | 0.66 | 4.17 | 4.50 | 16.41 | 2.63 | 0.00 | 3.02 |
| Road slope (event 8) | 0.19 | 4.94 | 3.57 | 0.87 | 0.64 | 4.04 | 4.74 | 23.57 | 9.80 | 0.00 | 2.71 |
| Road slope (event 9) | -0.96 | 7.02 | 2.45 | 5.19 | 1.24 | 7.81 | 2.24 | 44.10 | 2.68 | 0.00 | 2.43 |
| Road slope (event 10) | 0.29 | 8.83 | 2.04 | 1.68 | 1.44 | 9.06 | 2.00 | 6.56 | 1.25 | 0.04 | 1.18 |
| Roesgrenda | 0.38 | 6.45 | 2.65 | 2.50 | 1.03 | 6.48 | 2.64 | 6.80 | 1.47 | 0.03 | 1.43 |
| Ruinon | -1.4 | 13.60 | 1.59 | 9.37 | 2.15 | 13.49 | 1.59 | 48.30 | 13.18 | 0.00 | 2.66 |
| Séchilienne | 0.81 | 9.52 | 1.93 | 2.13 | 1.48 | 9.29 | 1.97 | 20.72 | 5.61 | 0.00 | 2.57 |
| Takabayama | -0.28 | 4.94 | 3.57 | 1.30 | 0.78 | 4.90 | 3.61 | 19.21 | 22.30 | 0.00 | 2.99 |
| Vajont (#2) | -0.98 | 5.59 | 3.08 | 10.48 | 0.80 | 5.04 | 3.48 | 15.89 | 1.25 | 0.00 | 1.29 |
| Vajont (#4) | 0.21 | 6.89 | 2.49 | 0.81 | 1.11 | 6.95 | 2.47 | 20.34 | 2.00 | 0.00 | 1.95 |

**Table S2 (continued).** Parameters of the LPPLS calibration and Lomb analysis for the different landslides.

| Site | $m_{LPPLS}$ | $\omega_{LPPLS}$ | $\lambda_{LPPLS}$ | $C/B$ (%) | $f_{Lomb}$ | $\omega_{Lomb}$ | $\lambda_{Lomb}$ | $P_{max}$ | $\eta$ | $p_{FA}$ | $\gamma$ |
|---|---|---|---|---|---|---|---|---|---|---|---|
| Vajont (#6) | 0.02 | 4.94 | 3.57 | 0.18 | 0.63 | 3.97 | 4.86 | 33.79 | 3.48 | 0.00 | 4.99 |
| Vajont (#58) | -0.025 | 4.94 | 3.57 | 0.14 | 0.66 | 4.18 | 4.50 | 41.40 | 13.79 | 0.00 | 3.20 |
| Veslemannen (2017; #1) | -0.86 | 7.87 | 2.22 | 8.99 | 1.22 | 7.65 | 2.27 | 18.12 | 5.90 | 0.00 | 2.88 |
| Veslemannen (2017; #2) | -0.43 | 6.68 | 2.56 | 3.82 | 1.03 | 6.50 | 2.63 | 18.40 | 8.19 | 0.00 | 3.76 |
| Veslemannen (2017; #3) | -0.71 | 7.83 | 2.23 | 5.30 | 1.17 | 7.38 | 2.34 | 18.24 | 6.27 | 0.00 | 3.12 |
| Veslemannen (2017; #4) | -0.71 | 7.76 | 2.25 | 5.67 | 1.23 | 7.74 | 2.25 | 17.80 | 7.38 | 0.00 | 3.10 |
| Veslemannen (2017; #5) | -0.96 | 7.39 | 2.34 | 10.75 | 1.23 | 7.73 | 2.25 | 13.08 | 11.01 | 0.00 | 1.71 |
| Veslemannen (2017; #6) | -0.42 | 7.73 | 2.25 | 2.93 | 1.18 | 7.41 | 2.33 | 17.53 | 6.62 | 0.00 | 3.44 |
| Veslemannen (2017; #7) | -0.5 | 6.92 | 2.48 | 3.66 | 1.06 | 6.67 | 2.56 | 18.43 | 9.79 | 0.00 | 3.46 |
| Veslemannen (2018; #1) | -0.25 | 4.94 | 3.57 | 1.62 | 0.76 | 4.77 | 3.73 | 23.49 | 4.42 | 0.00 | 1.94 |
| Veslemannen (2018; #2) | -0.021 | 4.94 | 3.57 | 0.15 | 0.77 | 4.85 | 3.65 | 18.70 | 2.53 | 0.00 | 2.12 |
| Veslemannen (2018; #3) | 0.022 | 5.00 | 3.51 | 0.17 | 0.79 | 4.94 | 3.57 | 21.24 | 3.58 | 0.00 | 2.72 |
| Veslemannen (2018; #4) | -0.31 | 5.98 | 2.86 | 2.24 | 0.99 | 6.22 | 2.75 | 23.87 | 5.47 | 0.00 | 2.59 |
| Veslemannen (2018; #5) | -0.12 | 5.13 | 3.40 | 1.58 | 0.79 | 4.94 | 3.57 | 24.44 | 8.29 | 0.00 | 4.37 |
| Veslemannen (2018; #6) | 0.29 | 4.95 | 3.56 | 2.68 | 0.79 | 4.97 | 3.54 | 23.48 | 5.78 | 0.00 | 3.65 |
| Veslemannen (2018; #7) | -0.11 | 5.71 | 3.01 | 1.53 | 0.89 | 5.61 | 3.06 | 24.06 | 10.28 | 0.00 | 4.98 |
| Veslemannen (2019; #1) | 0.67 | 6.42 | 2.66 | 9.37 | 0.96 | 6.06 | 2.82 | 40.31 | 1.20 | 0.00 | 2.52 |
| Veslemannen (2019; #2) | 0.47 | 7.52 | 2.31 | 4.17 | 1.26 | 7.92 | 2.21 | 31.27 | 6.84 | 0.00 | 2.36 |
| Veslemannen (2019; #3) | 0.43 | 6.51 | 2.63 | 4.62 | 1.06 | 6.66 | 2.57 | 37.45 | 5.08 | 0.00 | 2.35 |
| Veslemannen (2019; #4) | 0.45 | 6.43 | 2.66 | 5.33 | 1.06 | 6.64 | 2.58 | 32.99 | 1.26 | 0.00 | 1.89 |
| Veslemannen (2019; #5) | 0.55 | 6.46 | 2.64 | 6.31 | 1.07 | 6.70 | 2.55 | 25.58 | 1.45 | 0.00 | 1.46 |
| Veslemannen (2019; #6) | 0.28 | 6.59 | 2.59 | 2.64 | 1.03 | 6.48 | 2.64 | 30.27 | 1.14 | 0.00 | 1.87 |
| Veslemannen (2019; #7) | 0.52 | 6.57 | 2.60 | 6.07 | 1.07 | 6.71 | 2.55 | 23.07 | 1.39 | 0.00 | 1.33 |
| Welland (point 1) | -0.5 | 4.94 | 3.57 | 6.36 | 0.76 | 4.76 | 3.74 | 30.35 | 4.01 | 0.00 | 3.43 |
| Welland (point 2) | -0.47 | 4.96 | 3.55 | 7.53 | 0.81 | 5.09 | 3.43 | 39.63 | 3.57 | 0.00 | 2.08 |
| Xintan | -0.96 | 4.94 | 3.57 | 11.36 | 0.70 | 4.40 | 4.17 | 7.56 | 1.33 | 0.02 | 1.38 |

Note: $m_{LPPLS}$ is the power law exponent, $\omega_{LPPLS}$ is the angular log-periodic frequency, $\lambda_{LPPLS}$ is the scaling ratio, and $C/B$ is the relative log-periodic oscillation amplitude, which are all derived from the LPPLS model calibration; $f_{Lomb}$ is the log frequency, $\omega_{Lomb}$ is the angular log-periodic frequency, $\lambda_{Lomb}$ is the scaling ratio, $P_{max}$ is the Lomb peak height, $\eta$, is the first to second highest peak ratio, $p_{FA}$ is the false-alarm probability, and $\gamma$ is the signal-to-noise ratio, which are all derived from the Lomb periodogram analysis.